\documentclass[useAMS,usenatbib]{mn2e}

\usepackage{graphicx}
\usepackage{times}
\usepackage{amssymb,alltt,amsmath}
\usepackage{color}
\usepackage{hyperref}
\usepackage{supertabular}
\hypersetup{colorlinks, linkcolor=blue}

\title[The Extraordinary Abell 2744]{The Extraordinary Amount of Substructure in the \emph{Hubble Frontier Fields} Cluster Abell 2744}
\author[Jauzac et al. 2016]
{M. Jauzac,$^{1,2,3}$\thanks{E-mail:
mathilde.jauzac@durham.ac.uk} D. Eckert,$^4$ 
J. Schwinn,$^{2,5}$ D. Harvey,$^6$ C. M. Baugh,$^2$ A. Robertson,$^2$ 
\newauthor
S. Bose,$^{2}$ R. Massey,$^{1,2}$ M. Owers,$^{7,8}$ H. Ebeling,$^{9}$ H. Y. Shan,$^6$ E. Jullo,$^{10}$ J.-P. Kneib,$^{6}$  
\newauthor
J. Richard,$^{11}$ H. Atek,$^{12}$ B. Cl\'ement,$^{11}$  E. Egami,$^{13}$ H. Israel,$^{14,1,2}$ K. Knowles,$^{15}$   
\newauthor
M. Limousin,$^{7}$ P. Natarajan,$^{12}$ M. Rexroth$^{6}$, P. Taylor,$^{1}$ C. Tchernin$^5$
\\
\\
$^{1}$Centre for Extragalactic Astronomy, Department of Physics, Durham University, Durham DH1 3LE, U.K.\\
$^{2}$Institute for Computational Cosmology, Durham University, South Road, Durham DH1 3LE, U.K.\\
$^{3}$Astrophysics and Cosmology Research Unit, School of Mathematical Sciences, University of KwaZulu-Natal, Durban 4041, South Africa\\
$^{4}$Astronomy Department, University of Geneva, 16 ch. d'Ecogia, CH-1290 Versoix, Switzerland\\
$^{5}$Zentrum f\"ur Astronomie, Institut f\"{u}r Theoretische Astrophysik, Universit\"{a}t Heidelberg, Philosophenweg 12, D-69120 Heidelberg, Germany\\
$^{6}$Laboratoire d'Astrophysique, Ecole Polytechnique F\'ed\'erale de Lausanne (EPFL), Observatoire de Sauverny, CH-1290 Versoix, Switzerland\\
$^{7}$Department of Physics and Astronomy, Macquarie University, NSW 2109, Australia \\
$^{8}$Australian Astronomical Observatory PO Box 915, North Ryde NSW 1670, Australia \\
$^{9}$Institute for Astronomy, University of Hawaii, 2680 Woodlawn Drive, Honolulu, Hawaii 96822, USA\\
$^{10}$Laboratoire d'Astrophysique de Marseille - LAM, Universit\'e d'Aix-Marseille $\&$ CNRS, UMR7326, 38 rue F. Joliot-Curie, 13388 Marseille Cedex 13, France\\
$^{11}$Univ Lyon, Univ Lyon1, ENS de Lyon, CNRS, Centre de Recherche Astrophysique de Lyon UMR5574, F-69230, Saint-Genis-Laval, France\\
$^{12}$Department of Astronomy, Yale University, 260 Whitney Avenue, New Haven, CT 06511, USA\\
$^{13}$Steward Observatory, University of Arizona, 933 North Cherry Avenue, Tucson, AZ, 85721, USA \\
$^{14}$Faculty of Physics, Ludwig-Maximilians Universit\"{a}t, Scheinerstr.\ 1, 81679 Munich, Germany \\
$^{15}$Astrophysics and Cosmology Research Unit, School of Chemistry and Physics, University of KwaZulu-Natal, Durban 4041, South Africa
}

\begin{document}

\date{Accepted XXXX. Received 2016; in original form 2016 June 14}

\pagerange{\pageref{firstpage}--\pageref{lastpage}} \pubyear{2016}

\maketitle

\label{firstpage}

\begin{abstract}
We present a joint optical/X-ray analysis of the massive galaxy cluster Abell 2744 ($z$=0.308). Our strong- and weak-lensing analysis within the central region of the cluster, i.e., at $R<1$~Mpc from the brightest cluster galaxy, reveals eight substructures, including the main core. All of these dark-matter halos are detected with a significance of at least $ 5 \sigma$ and feature masses ranging from 0.5 to 1.4$\times$10$^{14}$M$_{\odot}$ within $R<150$~kpc. 
\citet{merten11} and \citet{medezinski16} substructures are also detected by us. We measure a slightly higher mass for the main core component than reported previously and attribute the discrepancy to the inclusion of our tightly constrained strong-lensing mass model built on \emph{Hubble Frontier Fields} data.
X-ray data obtained by \textit{XMM-Newton} reveal four remnant cores, one of them a new detection, and three shocks. Unlike \cite{merten11}, we find all cores to have both dark and luminous counterparts. 

A comparison with clusters of similar mass in the MXXL simulations yields no objects with as many massive substructures as observed in Abell 2744, confirming that Abell 2744 is an extreme system. We stress that these properties still do not constitute a challenge to $\Lambda$CDM, as caveats apply to both the simulation and the observations: for instance, the projected mass measurements from gravitational lensing and the limited resolution of the sub-haloes finders.

We discuss implications of Abell 2744 for the plausibility of different dark-matter candidates and, finally, measure a new upper limit on the self-interaction cross-section of dark matter of $\sigma_{\rm DM} < 1.28\, $cm$^2$g$^{-1}$(68\% CL), in good agreement with previous results from \citet{harvey15}. 
\end{abstract}



\section{Introduction}
\label{intro}
Galaxy clusters are the (so far) final stage in the evolution of cosmic large-scale structure \citep{bond96}.
Their present-day structure encodes a rich history of continuous accretion from their surroundings and occasional  mergers with other clusters \citep[][]{schaye15,vogelsberger14,springel06,evrard02,colless01,GH89}. As such,
clusters are ideal laboratories to study the mass assembly processes of the cosmic web.

85\% of the mass in galaxy clusters is invisible dark matter. The standard $\Lambda$CDM (Cold Dark Matter) theory posits that dark matter consists of non-relativistic, non-baryonic particles that interact with ordinary matter only via the force of gravity. However, discrepancies with observational evidence from studies of low-mass substructure have led to suggestions of Warm Dark Matter \citep[WDM; ][]{viel11}, Self-Interacting Dark Matter \citep[SIDM; ][]{spergel00}, interacting Dark Matter \citep[$\gamma$-DM, and $\nu$-DM; ][]{boehm14}. 
The difference between the properties of dark and ordinary matter becomes most strikingly apparent by dark matter's behaviour during collisions between components of a cluster's substructure \citep[][]{clowe04,bradac06,bradac08a,harvey15,massey15}, especially when extremely massive clusters are involved.

The most direct method to detect dark matter exploits gravitational lensing of background objects \citep[for reviews see ][]{massey10,KN11,hoekstra13}. A combination of weak (linear) lensing plus strong (non-linear) lensing techniques can reconstruct the distribution of total mass (luminous as well as dark) from the cluster's inner core to the outskirts and the connecting large-scale structure. Multi-wavelength data (e.g.\ X-ray imaging to trace the intra-cluster medium, and near-infrared photometry to measure stellar masses in cluster galaxies) lend crucial support to such studies by enabling us to deduce the dynamics of substructures within the cluster \citep[][]{bradac06,bradac08a,owers11,merten11,ogrean15,eckert15,jauzac15a} and hence constrain their evolution.

Abell 2744 (also known as AC118 and MACSJ0014.3-3022, $z=0.308$) is one of the most massive galaxy clusters known. Its highly disturbed dynamical state was investigated and emphasised by \citet{owers11} and \citet{merten11}, based on X-ray, dynamical, and lensing studies. Being a powerful gravitational lens, A2744 made an ideal target for the \emph{Hubble Frontier Fields} campaign\footnote{http://www.stsci.edu/hst/campaigns/frontier-fields/} \citep[HFF, ][]{lotz16} which obtained the deepest imaging data to date for galaxy clusters (mag$_{\rm AB, lim}\sim29$ in seven passbands from the optical to the near-infrared, corresponding to 140 orbits per cluster). The resulting data enabled a wide range of investigations into the properties of A2744 \citep[][]{atek14a,laporte14,wang15,jauzac15b,lam14,zheng14,montes14,zitrin14,rawle14,atek15,schirmer15,kawamata15,zitrin15,eckert15,merlin16,castellano16,eckert16,medezinski16,pearce16}. 

The work presented here complements our previous studies of A2744 \citep{jauzac15b,eckert15}. \cite{jauzac15b} focused on the inner core of Abell 2744 using the HFF strong-lensing data, while \cite{eckert15} explored the mass distribution on very large scales ($R\sim 4$~Mpc). The latter study combined strong and weak lensing just like our analysis here, but concentrated on the mass distribution along the three large-scale filaments detected with \emph{XMM-Newton}. The results presented here on the mass distribution within the central $1$~Mpc of A2744 thus connects the scales investigated by us in prior work, in an attempt to obtain a complete picture of the mass distribution in this exceptional cluster.

Our paper is organised as follows:
Sect.~\ref{observations} summarises the data used in this work. Sect.~\ref{SLanalysis} and Sect.~\ref{WLgal} describe our strong- and weak-lensing analysis, respectively. The mass modelling technique employed by us to combine strong- and weak-lensing constraints is explained in Sect.~\ref{slwl_model}; the resulting mass distribution, including numerous substructures, is presented in Sect.~\ref{lensing_mass}. Complementing our lensing study, Sect.~\ref{xray_thermo} presents insights gleaned from the X-ray emission of the diffuse intra-cluster gas, and Sect.~\ref{dynamics} provides a revised dynamical analysis of the cluster.  Finally, we compare our results with the $\Lambda$CDM MXXL simulation in Sect.~\ref{discussion} and also discuss implications of our findings for the nature of dark matter.

Throughout, we adopt the $\Lambda$CDM concordance cosmology model with $\Omega_{m}=0.3$, $\Omega_{\Lambda}=0.7$, and a Hubble constant $H_{0}=70$km.s$^{-1}$.Mpc$^{-1}$. Magnitudes are quoted in the AB system.

\section{Observations}
\label{observations}

\subsection{\textit{Hubble Space Telescope}}
Abell 2744 was imaged with the \textit{Hubble Space Telescope} (HST) during Cycle 17 as part of programme GO-11689, PI: R. Dupke). These observations consist of two tiles with $\sim$50\% overlap, taken with the \emph{Advanced Camera for Surveys} (ACS) in three filters (F435W, F606W, and F814W), and extending over four \textit{HST} orbits for each tile and in each filter. The resulting data were used in the first analysis of the dynamics of Abell 2744 published by \cite{merten11}, and later by several teams of researchers \citep{coe15,johnson14,richard14} in the context of the HFF mass mapping initiative in order to create the first mass models for release to the astronomical community.

More recently, the core of Abell 2744 was observed as part of the HFF initiative \citep[ID: 13495, PI: J. Lotz, ][]{lotz16} with the \textit{Wide-Field Camera 3} (WFC3) between October 25$^{\rm th}$ and November 28$^{\rm th}$ 2013 in four filters (F105W, F125W, F140W, and F160W) for total integration times of 24.5, 12, 10, and 14.5 orbits respectively. Additional observations with ACS were obtained seven months later, between May 14$^{\rm th}$ and July 1$^{\rm st}$ 2014, in three filters (F435W, F606W, and F814W) for total integration times of 24, 14, and 46 orbits, respectively. 

The strong-lensing model of Abell 2744 used by us here is based on the self-calibrated data (version v1.0) with a pixel size of 0.03$\arcsec$, provided by STScI\footnote{https://archive.stsci.edu/pub/hlsp/frontier/a2744/images/hst/} \citep[more details are given in][]{jauzac15b}. The \textit{HST} weak-lensing measurements relied on custom-reduced data adapted to shape measurements with multiple exposures, also provided by STScI\footnote{https://archive.stsci.edu/pub/hlsp/frontier/abell2744/images/hst/v1.0/ancillary}.

\subsection{\emph{Canada France Hawaii Telescope}}
Since the \textit{HST} data only probe the inner core of the cluster, we used  data from groundbased observations conducted with the \textit{Canada-France-Hawaii Telescope} (CFHT) to explore the mass distribution on larger scales.

Abell 2744 was imaged by the CFHT/MegaPrime in the \textit{i-}band on June 26$^{th}$ 2009 (PI: Martha Milkeraitis, 09AC24), in 10 exposures of 560s each, leading to a total integration time of 5.6~ks. The seeing was $\sim$0.91$\arcsec$.

The data were reduced using the public THELI pipeline \citep[][]{erben13}. THELI is a versatile image processing pipeline designed to handle optical and near-infrared imaging data from mosaic cameras.  In addition to bias subtraction, flat field correction, and the construction of pixel weight and flag maps, THELI performs astrometric and relative photometric calibration using \texttt{Scamp} \citep{bertin06} before producing a final coadded (stacked) image using \texttt{Swarp} \citep{bertin10}.  A common astrometric solution is thus established for all filters. THELI also includes an automated masking routine to identify bright stars and other image artefacts. We used the co-added, weighted mean stack to perform object detection and photometry.

\subsection{\emph{Wide Field Imager}}
Our photometric catalogue of galaxies in the Abell 2744 field is based on archival imaging obtained using the Wide Field Imager (WFI) on the MPG/ESO $2.2$~m telescope at La Silla Observatory, Chile. Abell 2744 was observed with WFI for multiple programmes between September 2000 and October 2011.
The peaks of the resulting number counts at 26.6, 26.5, 26.5, and 25.8\,mag in the R, V, B, and U passband, respectively, represent conservative estimates of the limiting magnitude in each filter, with a significant drop in number counts occurring only at $mag\approx 28$ in the R, V, and B bands, and at $mag\approx 27$ in the U band.
As for the CFHT data, the WFI data reduction in the $UBVR$ filters was performed using the THELI pipeline developed for WFI data.

In the $BVR$ filters, we use THELI coadded images created for a weak-lensing follow-up study (Klein et al.~(\textit{in prep.}) of clusters with Sunyaev-Zel'dovich effect observations with APEX-SZ; exposure times and seeing of these images are as follows: $B$: $9.2$~ks, $1\farcs87$ seeing; $V$: $8.7$~ks, $1\farcs51$ seeing; $R$: $21.0$~ks, $0\farcs87$ seeing.
The stacked image in the $U$ band created by us from the raw data; its total exposure time is $10.8$~ks, and the seeing is $1\farcs68$. The \texttt{automask} \citep{dietrich07} software was again employed to exclude image artefacts and saturated stars. 

Catalogues are generated for each passband using \texttt{SExtractor} \citep{BA96} in `dual image mode', with sources being detected in the deepest ($R$) image and their fluxes measured within the exact same aperture for all four filters. We do not apply PSF matching, but colour offsets due to seeing differences have been corrected using Stellar Locus Regression. All sources in the resulting catalogue are classified as either stars or galaxies based on their apparent $R$-band size and (un-calibrated) magnitude.

Stellar locus regression \citep[SLR]{high09}, in which the colour indices of measured stars are fitted against the known stellar locus, is used to calibrate the magnitude zeropoints in the $UBVR$ bands against one another, while also accounting for differences in seeing. Using the SLR zeropoint corrections thus determined, we compile the photometric catalogue including the positions, $UBVR$ magnitudes, and aperture ellipse parameters for $25000$ galaxies in the $34\arcmin\,\times 34\arcmin$ WFI field of view.

\subsection{\emph{XMM-Newton X-ray Observatory}}

A2744 was observed by \emph{XMM-Newton} on December 18--20, 2014 (OBSID 074385010, PI: Kneib) for a total exposure time of 110 ks. We reduced the data using the standard XMMSAS software package v13.5 and the ESAS data-reduction scheme \citep{snowden08}. Lightcurves were extracted for each of the three EPIC instruments and filtered to remove time periods of enhanced background caused by contamination from soft-proton flares. We used a collection of closed-filter observations to create a model image of the particle background. Long-term variations in the background rate were accounted for by rescaling the resulting image by the ratio between the count rates measured in the unexposed corners of the three instruments and the closed-filter observations. We then extracted source and background images in five energy bands (0.5--0.7, 0.7--1.2, 1.2--2.0, 2.0--4.0, and 4.0--7.0 keV). Vignetting effects were corrected by creating effective exposure maps in each band using the XMMSAS task \texttt{eexpmap}. Point sources detected by the XMMSAS task \texttt{ewavelet} were masked. To highlight regions of faint, diffuse signal, the image was adaptively smoothed using \texttt{asmooth} \citep{ebeling06}. For more details of the analysis procedure we refer to \citet{eckert15}.

\subsection{\emph{Chandra X-ray Observatory}}
\emph{Chandra} observed A2744 on several occasions \citep[2001-09-13 with ACIS-S for 25 ks, PI: David; 2006-11-08, 2007-06-10, and 2007-06-14, for a total of 100~ks with ACIS-I, PI: Kempner; see][]{owers11}. We analysed all archival data using CIAO v4.6 and the corresponding calibration database by reprocessing all individual observations using the CIAO task \texttt{chandra\_repro}, examining the light curves of each individual observation for flares, and filtering out periods of increased background. The resulting event files were then merged and a mosaic image extracted in the 0.7--7 keV band using the CIAO tool \texttt{fluximage}. 

\subsection{Spectroscopic Redshifts}
\label{zspec_obs}
In their combined X-ray / optical analysis of Abell 2744, \cite{owers11} used observations taken with the AAOmega multi-object spectrograph on the 3.9~m Anglo-Australian Telescope \citep[AAT; ][]{saunders04,smith04,sharp06}, during the nights of 12$^{th}$ to the 16$^{th}$ of September 2006. We here use Owers and co-workers' full catalogue of spectroscopic redshifts, which comprises 1237 objects lying within 15' of the cluster core, with 343 cluster members confirmed within 3~Mpc from the cluster centre. This catalogue combines the AAOmega spectra with those from the literature at the time of the analysis \citep[][]{boschin06,braglia09,couch98,couch87}. We note that \cite{boschin06} defined cluster membership as $c\, z_{cluster} \pm 4000\, {\rm km.s^{-1}}$, with $z_{cluster} = 0.308$, and refer the reader to \cite{owers11} for more details of these observations.

\begin{table}
\begin{center}
\begin{tabular}[h!]{cccc}
\hline
\hline
\noalign{\smallskip}
Component  & \#1 & \#2 & L$^*$ elliptical galaxy \\
\hline
$\Delta$ \textsc{ra}  & $-4.8^{+0.2}_{-0.1}$  &  $-15.5^{+0.1}_{-0.2}$ & --  \\
$\Delta$ \textsc{dec} & 4.0 $^{+0.1}_{-0.1}$  & $-17.0^{+0.2}_{-0.1}$ & --  \\
$e$ & 0.298 $\pm$0.004 & 0.595 $\pm$ 0.011 & -- \\
$\theta$ & 64.2$^{+0.3}_{-0.2}$  & 40.5$^{+0.4}_{-0.5}$ & -- \\
r$_{\mathrm{core}}$ (\footnotesize{kpc}) & 205.0$^{+1.1}_{-1.5}$  & 39.6$^{+0.8}_{-0.6}$  & [0.15] \\
r$_{\mathrm{cut}}$ (\footnotesize{kpc}) & [1000] & [1000] & 82.9$^{+6.7}_{-2.7}$ \\
$\sigma$ (\footnotesize{km\,s$^{-1}$}) &  1296$^{+3}_{-5}$ & 564$^{+2}_{-2}$ & 142$^{+5}_{-7}$ \\
\noalign{\smallskip}
\hline
\hline
\end{tabular}
\caption{\emph{Gold+Silver+Bronze} model best-fit PIEMD parameters for the two large-scale dark-matter halos, as well as for the L$^{*}$ elliptical galaxy. 
Coordinates are quoted in arcseconds with respect to $\alpha=3.586259, \delta=-30.400174$.
Error bars correspond to the $1\sigma$ confidence level. Parameters in brackets are not optimised.
The reference magnitude for scaling relations is $mag_{\rm{F814W}} = 19.44$.
}
\label{bestfit_SL}
\end{center}
\end{table}

\section{Strong-Lensing Analysis}
\label{SLanalysis}
We use an updated model from \cite{jauzac15b} for Abell 2744. Indeed, in the context of the 2015 \textit{Hubble Frontier Fields} Mass Mapping initiative, all lensing teams involved in the project shared data. For Abell 2744, spectroscopic redshifts of multiple images were given to the community by the GLASS team \citep[ID: 13459, PI: T. Treu;][]{schmidt14,treu15}, and were published in \cite{wang15}.
Through this process all teams voted for the different identified multiple images as \emph{`Gold'}, \emph{`Silver'} or \emph{`Bronze'} depending on how secure they think the systems/images were.

Using the multiple images identified as \emph{`Gold'}, \emph{`Silver'}, and \emph{`Bronze'}, i.e. 113 multiple images amongst 39 systems, we built a new mass model using the \textsc{Lenstool} software \citep[][]{jullo07}, following the methodology of \cite{jauzac15b}. The global cluster mass distribution is represented by two cluster-scale halos, as well as 733 galaxy-scale halos to include small-scale perturbations.
The mass model was run in the \emph{image plane}, and its \emph{best-fit} model predicts image positions with an RMS of 0.70$\arcsec$. The \emph{best-fit} parameters are given in Table~\ref{bestfit_SL}. This RMS value represents a slight improvement compared to the \cite{jauzac15b} model, for which we measured a global RMS over 157 images of 0.79\arcsec.
With this model, we measure a two-dimensional mass of M$(< 250\, {\rm kpc}) = 2.762\pm 0.006\, \times\, $10$^{14}\,$M$_{\odot}$.
This model is available to the community and can be found on the MAST\footnote{https://archive.stsci.edu/pub/hlsp/frontier/abell2744/models/cats/v3.1/}.

\section{Weak-Lensing Constraints}
\label{WLgal}

\subsection{\textit{HST} Weak Lensing Catalogue}
\subsubsection{The ACS Source Catalogue}
For the HST weak-lensing catalogue, the shape measurements are made in the ACS/F814W band. 
The methodology used to build the catalogue was presented in previous analyses \citep[][J12 and J15 hereafter]{jauzac12,jauzac15a}. We thus here only give a suummary of the procedure, and refer the reader to these papers for more details.
Our method is based on \cite{leauthaud07} (hereafter L07) who presented a weak-lensing analysis for the COSMOS survey.
We use the SE\textsc{xtractor} photometry package \citep{BA96} for the detection of the sources with the \textit{`Hot-Cold'} method \citep[][L07]{rix04}. 
The source catalogue is then cleaned by removing spurious or duplicate detections using a semi-automatic algorithm that identifies stars and saturated pixels, and designs polygonal masks around them. 
The galaxy-star separation is done by examining the distribution of objects in the magnitude (MAG$\_$AUTO) vs peak surface brightness (MU$\_$MAX) plane (see L07 $\&$ J12 for more details). Finally, the pattern-dependent correlations introduced by the drizzling process between neighboring pixels, which artificially reduces the noise level of co-added drizzled images, is corrected with care while applying the same remedy as L07 by simply scaling up the noise level in each pixel by the same constant F$_{A} \approx$ 0.316, defined by \cite{casertano00}. The resulting catalogue comprises 4582 sources identified as galaxies and 72 sources identified as point sources (stars) within a magnitude limit of $m_{\rm F814W} = 29.5$. 

One of the main steps in the build-up of a weak-lensing catalogue is the estimation and reduction of the contamination by non-lensed objects, i.e. cluster and foreground galaxies that would remain in the sample due to colours similar to background lensed galaxies. Their presence dilutes the observed shear and thus reduces the significance level of all quantities derived from it. Thus identifying and eliminating these contaminants represents a crucial step.
Cluster galaxies were identified thanks to the spectroscopic redshift catalogues published by \cite{owers11} containing 1237 objects, mentioned in Sect.~\ref{zspec_obs}. We also used photometric redshifts, derived by D. Coe (that were made available to the community in the context of the HFF mass mapping initiatives)\footnote{http://archive.stsci.edu/pub/hlsp/frontier/a2744/catalogues/hst/}. While this paper was being written, the \textsc{astrodeep} catalogues were made public \citep[][]{merlin16,castellano16}. We obtain similar results to the one presented here.
Following J12, the spectroscopic cluster membership criterion is defined by
$$z_{\rm cluster} - dz < z < z_{\rm cluster}+ dz ,$$
where $z$ is the spectroscopic redshift of the considered galaxy, $z_{\rm cluster} = 0.308$ is the systemic redshift of the cluster, and $dz = 0.0104$ is the 3$\sigma$ cut defined by the colour-magnitude selections presented in Sect.~\ref{CM_model}. 
Only 31\% of the sources in our ACS object catalogue have a photometric redshift. Of these 31\%, 20\% are identified as cluster members or foreground sources following the aforementioned selection criteria. 
To complement these redshift identifications, and identify contamination in the remaining sample of galaxies, we use a colour-colour diagram using the three HST/ACS bands to identify the regions dominated by foreground and cluster galaxies in the ($m_{\rm F435W}-m_{\rm F814W}$) -- ($m_{\rm F435W}-m_{\rm F606W}$) space. Its boundaries are calibrated by the spectroscopic and photometric redshifts, and defined as : 
$$m_{\rm F435W}-m_{\rm F814W} < 0.67776\, (m_{\rm F435W}-m_{\rm F606W}) + 0.3;$$ 
$$m_{\rm F435W}-m_{\rm F814W} > 0.87776\, (m_{\rm F435W}-m_{\rm F606W}) - 0.76;$$ 
$$m_{\rm F435W}-m_{\rm F814W} > 0.3$$

\begin{figure}
\hspace*{-3mm}\includegraphics[width=0.5\textwidth]{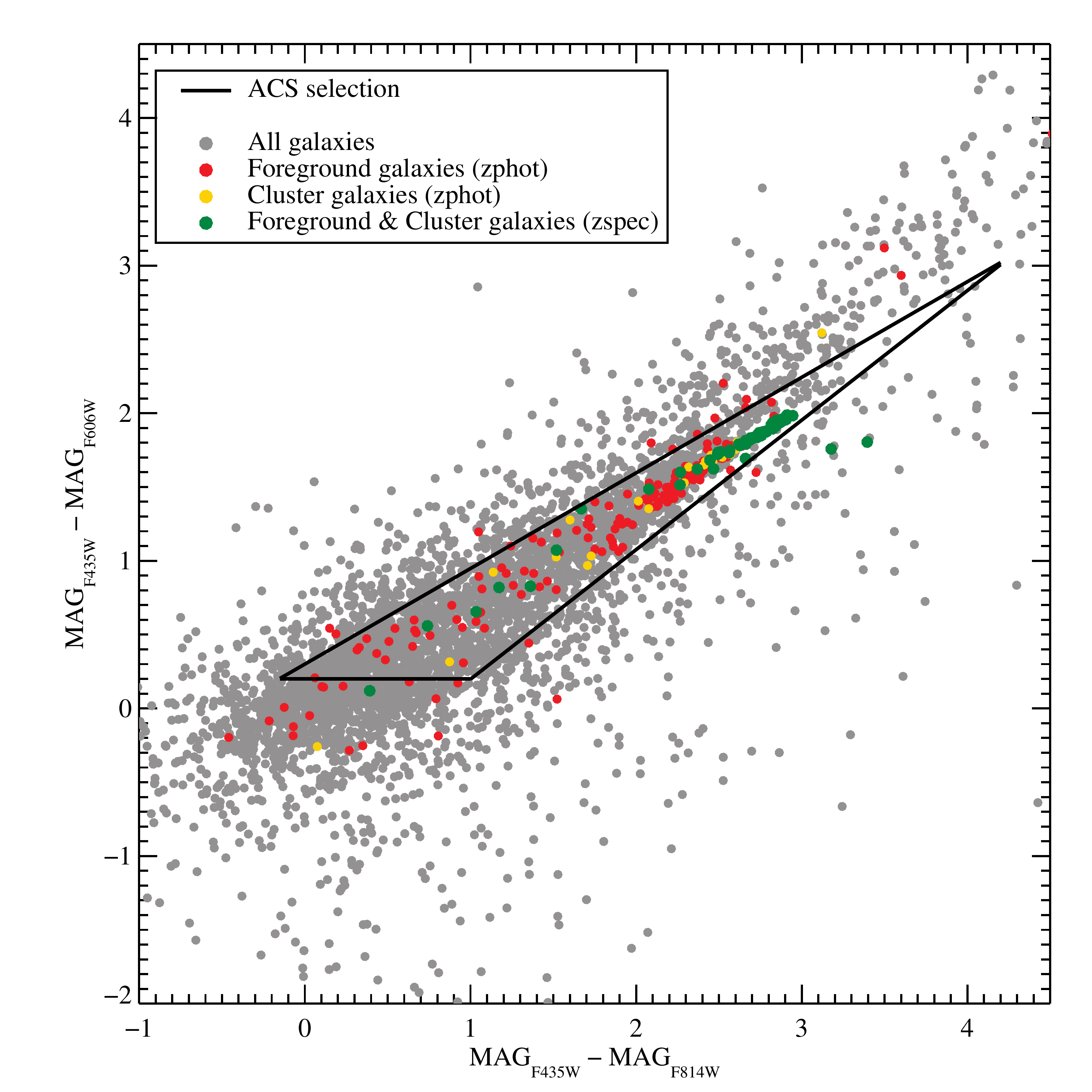}
\caption{Colour-colour diagram ($m_{\rm F435W}-m_{\rm F814W}$) vs ($m_{\rm F435W} - m_{\rm F606W}$) for objects in the HFF/ACS image of Abell 2744. Grey dots mark all galaxies in the study area. 
Unlensed galaxies diluting the shear signal are divided into several categories: galaxies spectroscopically confirmed as cluster members or foreground galaxies are marked in green, while galaxies classified as foreground objects (cluster members) via photometric redshifts are marked in red, and galaxies identified as cluster members via photometric redshifts are marked in yellow. The solid black lines delineate the colour-cut defined for this work to mitigate shear dilution by unlensed galaxies.
}
\label{cc_diagram}
\end{figure}

\begin{figure}
\hspace*{-3mm}\includegraphics[width=0.5\textwidth]{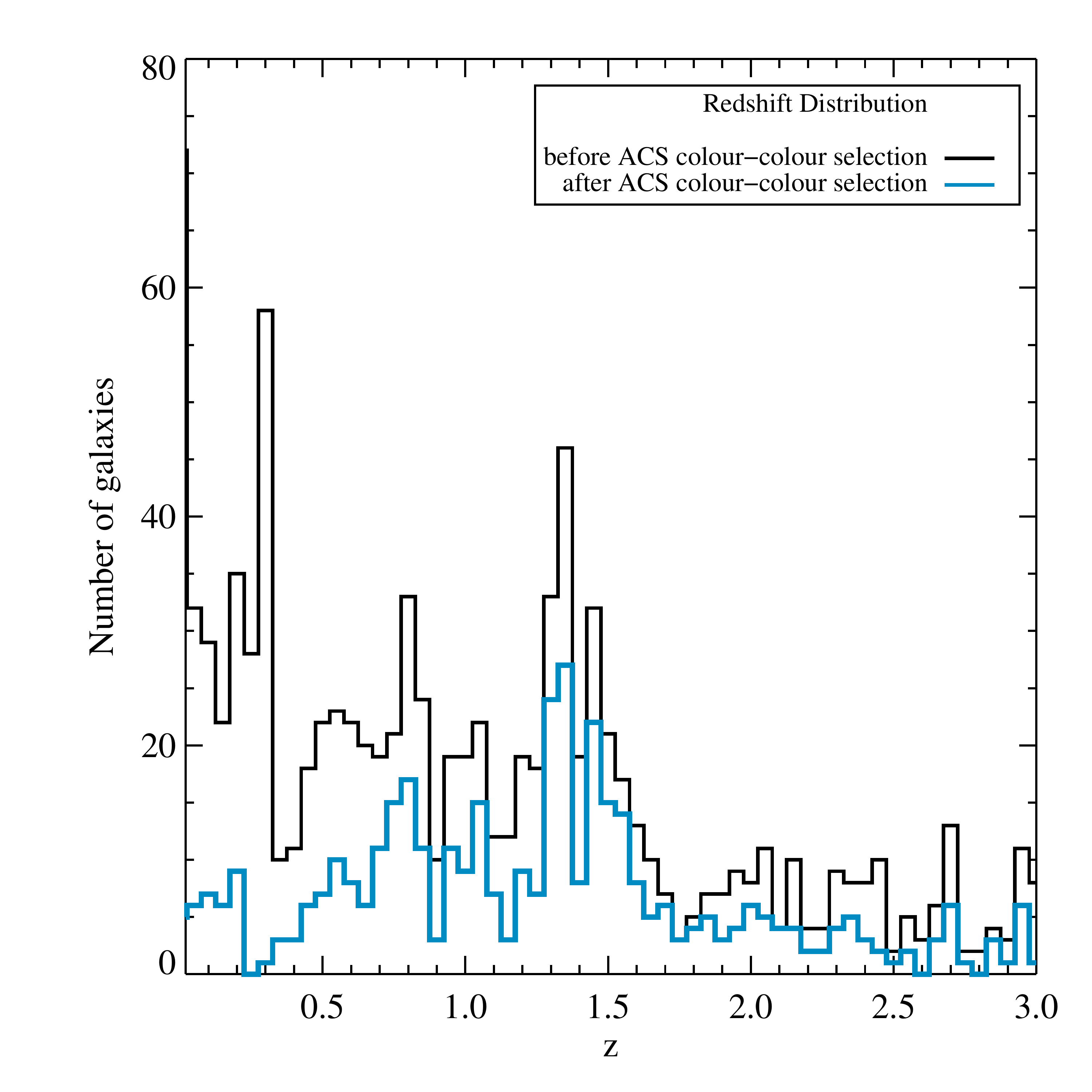}
\caption{
Redshift distribution of all galaxies with $m_{\rm F435W}$, $m_{\rm F606W}$, and $m_{\rm F814W}$ photometry from HFF observations. The black histogram corresponds to galaxies with photometric or spectroscopic redshifts; the cyan histogram is for galaxies classified as background objects using the colour-colour criterion illustrated in Fig.~\ref{cc_diagram}.
}
\label{Nz_ccselect}
\end{figure}

\begin{figure}
\hspace*{-3mm}\includegraphics[width=0.5\textwidth]{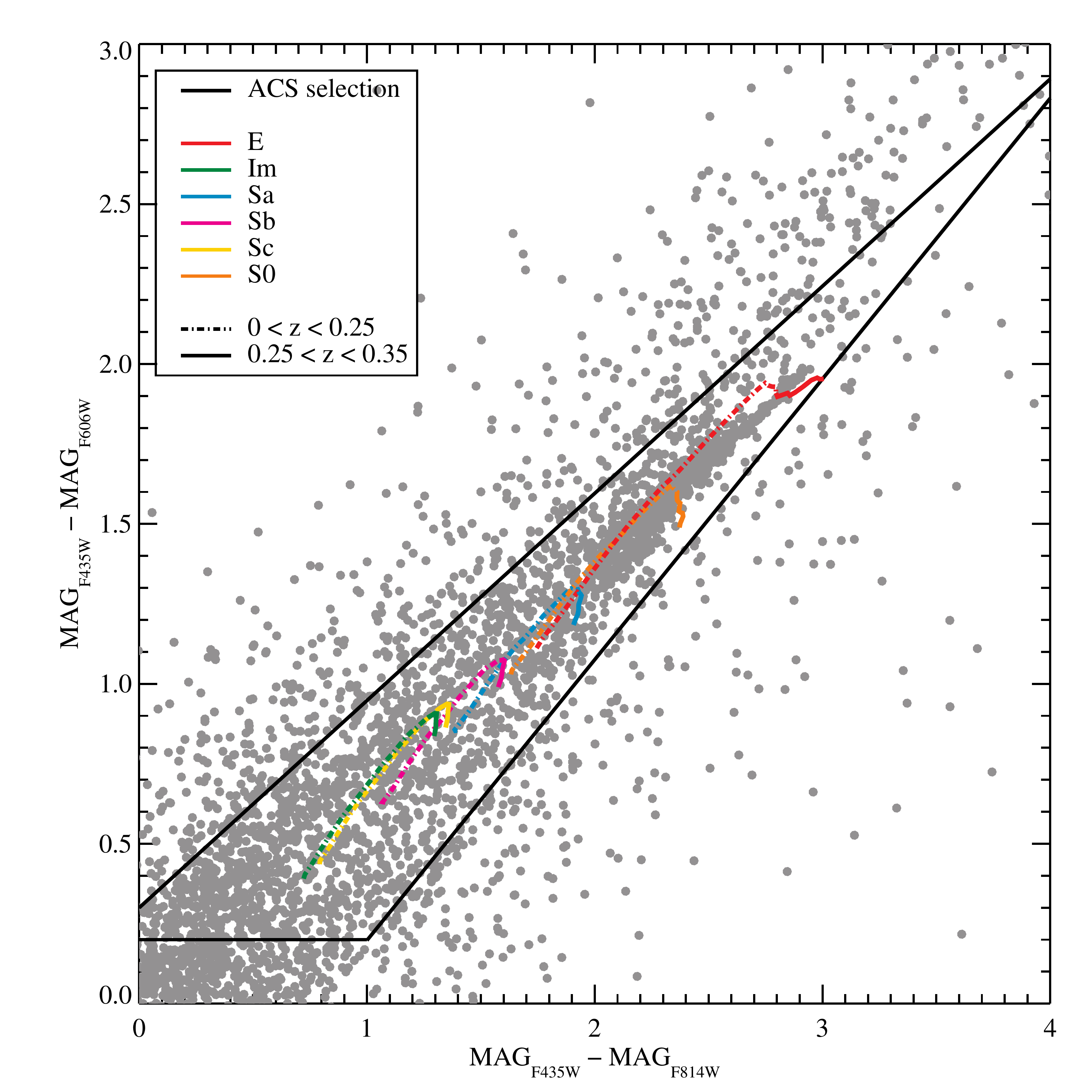}
\caption{
Colour-colour diagram ($mag_{F435W}-mag_{F814W}$) vs ($mag_{F435W} - mag_{F606W}$) as in Fig.~\ref{cc_diagram}. 
The solid black lines delineate the colour-cut defined for this work. The different spectral templates from \citet{CWW} and \citet{kinney96} as well as theoretical models from \citet{BC03} are marked by different colours: elliptical galaxies (red); Magellanic irregulars (green); spiral Sa galaxies (cyan); spiral Sb (magenta); spiral Sc (yellow); and S0 galaxies (orange). The dash--dotted curves correspond to a redshift range $0<z<0.25$ (foreground galaxies), the solid ones to $0.25<z<0.35$ (cluster members).
}
\label{cctracks}
\end{figure}

All objects within this region are removed from our analysis as presented in Fig.~\ref{cc_diagram}.
Fig.~\ref{Nz_ccselect} shows the galaxy redshift distribution before and after this F435W-F606W-F814W colour-colour selection. This selection is very efficient at removing cluster members and foreground galaxies at $z\leqslant$0.35 --- for the subset of our galaxies that have redshifts, 88\% of the unlensed population is eliminated.
As in J15, the validation of our colour-colour selection is done by predicting the colours expected from spectral templates at the redshift of the cluster or in the foreground. For this purpose, empirical templates from \citet{CWW} and \citet{kinney96} as well as theoretical templates from \cite{BC03} for various galaxy types in the Hubble sequences (ranging from Elliptical to SB) and starburst galaxies are used. The location of the colour-colour tracks at $z<0.35$ agree well with our selection region as shown in Fig.~\ref{cctracks} for the Bruzal \& Charlot model.

\subsubsection{Shape Measurements \& Lensing Cuts}
As in J12 and J15, we use the RRG method \citep{rhodes00} to measure the shapes of our background galaxy sample. RRG was specifically developed for space data. Later on \cite{rhodes07} adapted it to HST/ACS, to correct for the instability of the point-spread function of the instrument (PSF) over time-scales of weeks due to telescope `breathing'. This effect induces deviation from nominal focus, and the PSF thus becomes larger and more elliptical. To overcome this problem, a grid of simulated PSFs is created, and the effective focus of the observation is then determined by comparing the models with the ellipticity of $\sim$20 stars in each image. Following the \cite{bacon03} method, PSF parameters are then interpolated.
With the HFF data, we face another problem which is the time-scale between all exposures (observations were taken over several months), leading to a strong variation of the PSF pattern.
Therefore we model the PSF at each epoch, following the \cite{harvey15} RRG update, to obtain a more accurate estimation of the correction to apply to shear estimations. This method was proven to be successful in J15.

\cite{harvey15} adapted the RRG pipeline (L07) to model
the average PSF at the position of each galaxy in the stacked image. 
Using the identification of the stars positions from our initial ACS catalogue, using magnitude -- size and magnitude -- MU\_MAX diagrams, we measure their second and fourth moments  from {\em each} exposure and compare them to the ray tracing programme \textsc{TinyTim} model for the F814W band.
The PSF is then interpolated to the galaxy positions, the moments are rotated to be in the reference frame of the drizzled image, and then we average over the stacks. The PSF model then becomes dependent on the number of exposures covering the same area, and is thus not a continuous function across the field of view. Shear estimates done with fewer than 3 exposures are discarded, eliminating automatically edge galaxies and galaxies located near chip boundaries.

As presented in L07, the RRG output consist of three parameters: $d$, a measure of the galaxy size, as well as $e_1$ and $e_2$, the two components of the ellipticity vector $e = (e_1, e_2)$, defined as 
$$d = \sqrt{\frac{1}{2} (a^{2} + b^{2})},$$
$$e = \frac{a^{2} - b^{2}}{a^{2} + b^{2}},$$
$$e_{1} = e\, cos(2\phi),$$
$$e_{2} = e\,  sin(2\phi),$$
where $a$ and $b$ are respectively the major and minor axes of the background galaxy, and $\phi$ is the orientation angle of the major axis. Following L07, the ellipticity $e$ is calibrated by the shear polarisability, $G$, to obtain the shear estimator $\tilde{\gamma}$:
\begin{equation}
\tilde{\gamma} = C \frac{e}{G}\, ,
\label{eqn:shear}
\end{equation}
with the same global measurement as in L07:
$$G = 1.125 + 0.04 \arctan\frac{S/N - 17}{4}\, .$$
In Eq.\ref{eqn:shear}, $C$ is the calibration factor, derived from a set of simulated images similar to those used by STEP \citep{heymans06, massey07} for COSMOS images, and given by $C = (0.86^{+0.07}_{-0.05})^{-1}$.

The final step in constructing our weak-lensing catalogues consists of removing galaxies whose shape parameters are ill-determined, creating noise in shear measurements larger than the shear signal itself. We thus apply lensing cuts, following the ones presented in J12  and J15:
\begin{itemize}
\item
Threshold in the estimated detection significance:
$$\frac{S}{N} = \frac{FLUX\_AUTO}{FLUXERR\_AUTO} > 4.5 ;$$
\item
Threshold in the total ellipticity:
$$e = \sqrt{e_{1}^{2} + e_{2}^{2}} < 1 ;$$
\item
Threshold in the size, as defined by the RRG $d$ parameter :
$$3.6 < d < 30~\textrm{pixels}.$$
\end{itemize}

The RRG method allows ellipticities to be greater than 1 due to noise. We thus by definition, restrict the ellipticity to be $e \leqslant 1$. The limits applied to the size of the galaxies aim at eliminating: \textit{i)} small galaxies, theoretically smaller than the PSF itself, which thus would have non-accurate shape measurements, and \textit{ii)} large galaxies with a size similar to elliptical cluster members.
Finally, in order to ensure we are only considering galaxies that are weakly-lensed, we remove all galaxies within the multiple-image regions (where the non-linear regime dominates), which can be approximated by an ellipse aligned with the elongation of the cluster predicted by the strong-lensing model ($a=60\arcsec$, $b=42\arcsec$, $\theta = 60\deg$, $\alpha=3.5890837$ deg, $\delta=-30.399917$ deg).

The HFF-based catalogue extends to ACS-F814W magnitudes of 29, two magnitudes fainter than the pre-HFF dataset, which covers an extended region of the cluster \citep[as in][]{merten11}. 
We combined this new HFF catalogue with our pre-HFF one for the region not covered by HFF \citep[details given in][]{richard14}.
In the HFF region, more stars are saturated, thus the corresponding masks have to be increased; in total $\sim$45\% of the total (HFF+preHFF) ACS surface is masked out. Our final weak-lensing catalogue is composed of 1408 background galaxies, corresponding to a density of $\sim$ 110~galaxies.arcmin$^{-2}$. Compared to the catalogue generated by \cite{richard14}, presenting a pre-HFF weak-lensing catalogue, the density of weakly lensed galaxies has almost doubled.

\subsection{CFHT Weak Lensing Catalogue}
In order to extend the field of view over which we can probe the mass distribution, we combine our \textit{HST} analysis with CFHT weak-lensing data. We here summarize the procedure to build the background-galaxy catalogue from the CFHT data, we refer the reader to \cite{eckert15} for more details.
The sources are detected using \textsc{SExtractor} \citep[][]{BA96} in its standard mode.

To measure the galaxy shapes, we employed the popular KSB method \citep[][]{KSB95,LK97,hoekstra98}.
Our implementation of KSB is based on the KSBf90 pipeline\footnote{http://www.roe.ac.uk/~heymans/KSBf90/Home.html} as in \cite{shan12}.
For the PSF modeling, we identify stars in the size-MAG\_AUTO and MU\_MAX-MAG\_AUTO 
planes chip by chip as in \cite{shan12}. We then measure the Gaussian-weighted shape 
moments of the stars, and construct their ellipticity. In addition to cuts in MU\_MAX
and magnitude, we also exclude noisy outliers with signal-to-noise $S/N<100$ or absolute 
ellipticity more than 2$\sigma$ away from the mean local value, which can help us to iteratively 
remove objects very different from neighboring stars. Having obtained our clean 
sample of stars, a second order polynomial model in (x, y) is used to model the PSF 
across the field of view.  

Background galaxies are then selected following similar criteria to the ones applied to the \textit{HST} weak-lensing catalogue:
\begin{itemize}
\item Threshold in the magnitude distribution :
$$20<mag_{i}<26 ; $$
\item Threshold in the size :
$$ 1.15\, r_{PSF} < r_{h} < 10.0\, {\rm pixels} ; $$
\item Threshold in the estimated detection significance :
$$ \frac{S}{N} > 10 ; $$
\end{itemize}
where r$_{h}$ is the half-light radius, and $r_{PSF}$ is the size of the
largest star. Finally we apply a selection on the \textsc{Sextractor} parameter $FLAGS$, only keeping objects with $FLAGS=0$.
Our final CHFT weak-lensing catalogue has a background galaxy density of $\sim$10 galaxies.arcmin$^{-2}$.

\section{Strong \& Weak Lensing Mass Modelling}
\label{slwl_model}

\subsection{Combining Strong \& Weak Lensing Constraints}
\label{lenstool_grid}
In J15, we presented a new modeling approach implemented in \textsc{Lenstool} to combine both strong- and weak-lensing constraints to obtain a global mass distribution of the studied cluster.
We here give a summary of the methodology employed, and refer the reader to J15 for more details.

To model the strong-lensing region, we use the best-fit mass model described in Sect.~\ref{SLanalysis}, and combine it with a set of Radial Basis Functions (RBFs) located at the nodes of a multi-scale grid which covers an extended region where weak-lensing dominates. We finally add the dPIE (\emph{dual Pseudo-Isothermal Elliptical}) potentials \citep[][]{eliasdottir07} to account for cluster members. This technique enables us to weigh the strong-lensing constraints properly and to not account for them twice in the model.
As presented previously, the SL region is parametrically modeled using 2 cluster-scale halos and 733 galaxy-scale halos over the HST field of view. As we are extending our analysis out to $\sim$2~Mpc from the cluster center, we add 916 galaxy-scale halos, identified using a WFI colour-magnitude selection \citep[for more details on the selection see][]{eckert15}.
To this parametric model, we add a uniform grid comprising 3122 RBFs. Each RBF is modeled by a dPIE potential \citep[][]{eliasdottir07}, and is fixed in position and size, only its amplitude is left as a free parameter. As described in \cite{jullo09}, the potential core radius $s$ is set to the distance between the RBF and its closest neighbour, and its cut radius $t$ is three times the core radius.


After different tests on the grid resolution, we obtained an optimum solution with a multi-scale grid of 3122 RBFs, with a separation $s = 16.23\arcsec$ (more details are given in Sect.~\ref{global-mass}). We remove all RBFs located in the center of the cluster, where the strong-lensing regime dominates and where we model the main cluster components using cluster-scale halos as described in Sect.~\ref{SLanalysis}.

The contribution from the two components of our model is summed to the observed ellipticity following this equation:

\begin{equation}
\label{eq:shearmat}
\bold{e_m} = M_{\gamma v} \bold{v} + \bold{e_{\rm param}} + \bold{n}\;,
\end{equation}

\noindent where vector $\bold{v}$ contains the amplitudes of the 3122 RBFs, vector $\bold{e_m} = [\bold{e_1}, \bold{e_2}]$ contains the individual shape measurements of the weak-lensing sources, and $\bold{e_{\rm param}}$ is the fixed ellipticity contribution from the parametric model. 
The intrinsic ellipticity and noise in our shape measurements are represented by $\bold{n}$, also called the Gaussian noise in the shape measurements. $M_{\gamma v}$ is the matrix containing the cross-contribution of each individual RBF to each individual weak-lensing source. Shear components are scaled by distance ratios between each individual source $S$, the cluster $L$, and the observer $O$. $M_{\gamma v}$ components are thus given by :

\begin{eqnarray}
\label{eq:dshear1}
\Delta_{1}^{(j,i)} &= &\frac{D_{LSi}}{D_{OSi}}\ \Gamma_{1}^i(|| \theta_i - \theta_j ||,\ s_i,\ t_i) \, , \\
\Delta_{2}^{(j,i)} &= &\frac{D_{LSi}}{D_{OSi}}\ \Gamma_{2}^i(|| \theta_i - \theta_j ||,\ s_i,\ t_i) \, ,
\end{eqnarray}

\noindent  with analytical expressions for $\Gamma_1$ and $\Gamma_2$ are given in \citet[][Equation\ A8]{eliasdottir09}. 
Cluster shear can be large, thus the assumption from Eq.~\ref{eq:dshear1} may not be strictly valid. However, the dominant lensing signal is traced by the parametric model, while the grid-based model contribution originates primarily from the weak-lensing regime where this assumption is sensible.

\subsection{Modeling of Cluster Members}
\label{CM_model}
As explained in Sect.~\ref{lenstool_grid}, 1649 cluster members are added to complement our grid of RBFs modelled as dPIE potentials.
Two complementary methods are used to select these galaxies. For the HST field of view, we apply the identification method described in \cite{richard14}, which is based on a double colour-magnitude selection. All galaxies that fall within 3$\sigma$ of a linear model of the cluster red sequence in both the ($m_{\rm F606W} - m_{\rm F814W}$) vs $m_{\rm F814W}$ and ($m_{\rm F435W} - m_{\rm F606W}$) vs $m_{\rm F814W}$ colour-magnitude diagrams, are considered as cluster members.
For the extended field of view, we use the same methodology and criterion, but applied to the WFI B, V and R-bands. Our final catalogue comprises 1649 galaxies.

In the model these galaxies act as small-scale perturbers, with their cut radius and velocity dispersions fixed and scaled from their luminosities in the HST/ACS F814W-band for the HST selected ones, and in the WFI/R-band for the WFI selected ones. This methodology was successfully used in previous analysis from our group, and was recently validated by \cite{harvey16}.
We derive L$^*$ in our filters of observation based on the K$^*$ magnitudes obtained by \cite{lin06} as a function of cluster redshift. Cut radius and velocity dispersion are then scaled relative to an $m_{K}^{*} = 19.76$ galaxy with velocity dispersion $\sigma^* = (119\pm20)$ km s$^{-1}$ and cut radius $r_{cut}^*  = (85\pm20)$ kpc for all galaxies in our catalogue.

\subsection{Priors and MCMC sampling}
As presented in our previous analysis, the parameter space is sampled with the MassInf algorithm implemented in the Bayesis library \citep{skilling98}, which \cite{jullo14} implemented in \textsc{lenstool}.
At each iteration, using the Gibbs approach, the most significant RBFs are identified and their amplitude is adjusted to fit the ellipticity measurements.
One prior of MassInf is the number of significant RBFs, however \cite{jullo14} demonstrated that it does not have a significant impact on the reconstruction. Thus the initial number of significant RBFs is set to 2\%, and the algorithm converges to about 4\%.

As in previous works from our team, a standard likelihood function is chosen as the objective function, which is assumed to have Gaussian noise. The MCMC optimization returns a large number of samples from which we can then estimate mean values and errors on the interesting quantities such as the mass density field and the amplification field.


\begin{figure*}
\includegraphics[width=\textwidth]{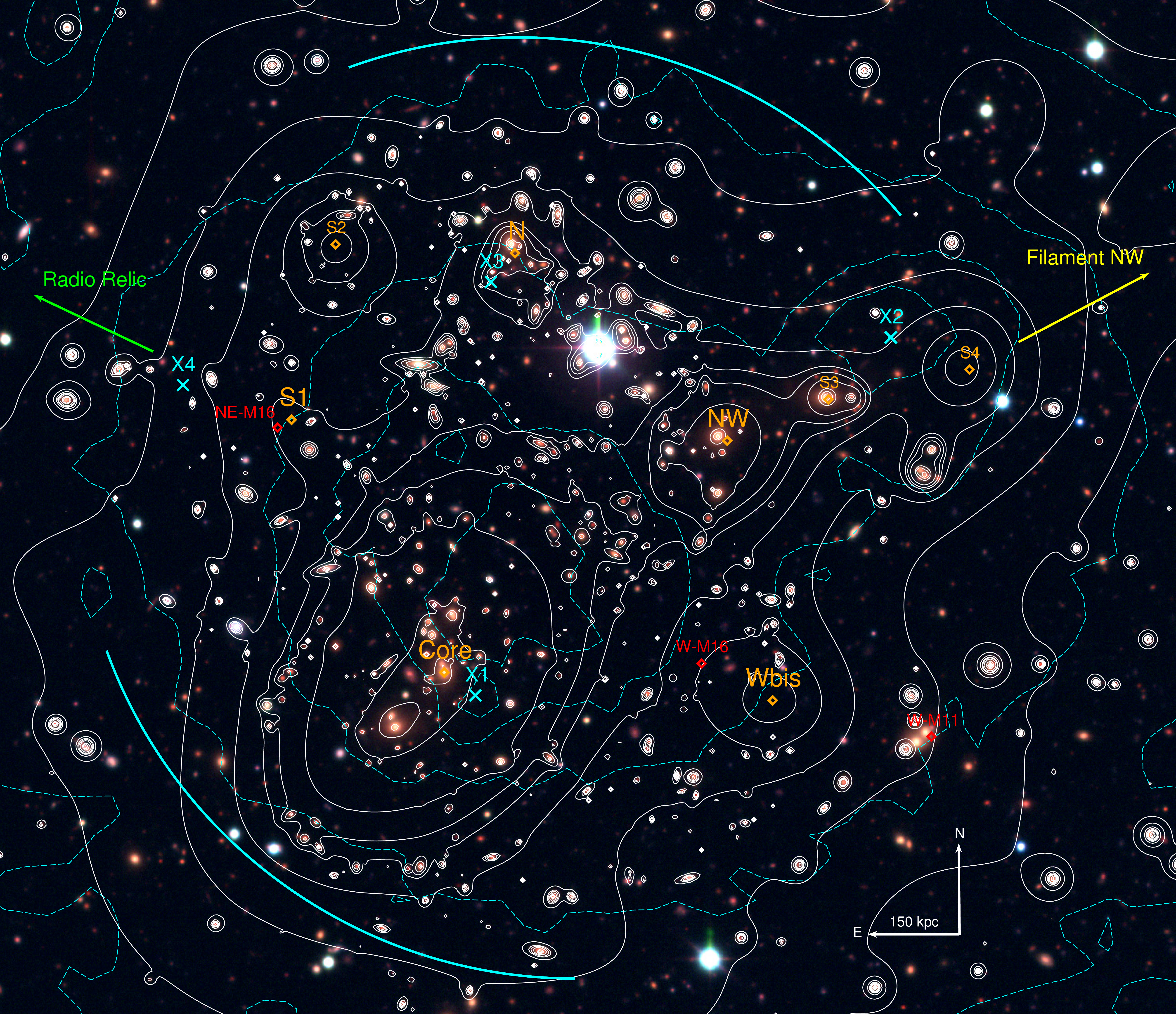}
\caption{
WFI Composite colour image of Abell 2744. Orange diamonds highlight the position of substructures detected in the strong+weak lensing mass map (and listed in Table~\ref{tab_substr}); red diamonds mark the position of the \emph{West} clump as in \citet{merten11} and as in \citet{medezinski16}, as well as the NE clump as in \citet{medezinski16}; and cyan crosses highlight the positions of remnant cores detected in the \emph{Chandra} and \emph{XMM-Newton} maps. White contours show the mass distribution derived from our strong+weak lensing mass model; cyan contours represent the gas distribution deduced from \emph{Chandra} observations. Cyan arcs highlight the position of the two shocks detected in X-rays and discussed in this paper. The yellow arrow highlights the direction of the NW filament first reported by \citet{eckert15}, while the green arrow denotes the direction in which the radio relic discussed in \citet{eckert16} is found.
}
\label{a2744_colour}
\end{figure*}

\subsection{Redshift Estimation for Background Sources}
Our background galaxy catalogue contains 7546 objects, 546 of which have a known redshift. For the 7000 remaining, we apply the following function to provide a good description of the redshift distribution of these background galaxies :
\begin{equation}
    \mathcal{N}(z) \propto e^{- (z /z_0)^\beta} \, ,
\end{equation}
\noindent with $\beta = 1.84$ and a median redshift $<z>= 1.586 = 0.56\, z_0$ \citep[][]{gilmore09,natarajan97}.

This method has proven to be successful in J15.
In addition, we split the catalog into a bright and a faint subsample at the median magnitude $m_{\rm F814W} = 26.4$.
Within the uncertainties given by the number statistics, the resulting two histograms have the same slope.
Since \textsc{Lenstool} allows each source to have its own redshift, we randomly draw (during the initialization phase) redshifts from the fitted redshift distribution for all  galaxies without spectroscopic or photometric redshift.

\section{Abell 2744 Mass Distribution}
\label{lensing_mass}

\begin{table*}
\begin{center}
\begin{tabular}[h!]{ccccccc}
\hline
\hline
\noalign{\smallskip}
$ID$ & R.A.\ (deg)& Dec.\ (deg) & $M (10^{13}\, M_{\odot})$ & $\sigma$ & $M/L$ & $D_{C-S}$ (kpc) \\
\hline
\emph{Core} & 3.586259 & -30.400174 & $13.55 \pm 0.09$ & 150 & 85 & -- \\
\emph{N} & 3.5766583 & -30.357592 & $6.10 \pm 0.50$ & 12 & 60 & 708.4 \\
\emph{NW} & 3.5530963 & -30.376764 & $7.90 \pm 0.60$ & 13 & 46 & 603.6\\
$W_{bis}$ & 3.5462875 & -30.403319 & $5.20 \pm 0.60$ & 9 & $>100$ & 565.3 \\
S1 & 3.6041246 & -30.37465 & $5.00 \pm 0.40$ & 13 & $>100$ & 486.9 \\
S2 & 3.59895 & -30.356925 & $5.40 \pm 0.50$ & 11 & $>100$ & 728.5 \\
S3 & 3.5415083 & -30.373778 & $6.50 \pm 0.60$ &11 & 51 & 763.7 \\
S4 & 3.524725 & -30.369583 & $5.50 \pm 1.20$ & 5 & $>100$ & 1000.5\\
\noalign{\smallskip}
\hline
\hline
\end{tabular}
\caption{Coordinates, mass within a 150\, kpc aperture, significance of detection, Mass-to-Light ratio (M/L) and distance to the cluster centre ($D_{C-S}$) for the substructures detected in the field of Abell 2744. 
The $W_{bis}$ substructure is consistent with \citet{medezinski16} \emph{W} substructure. 
}
\label{tab_substr}
\end{center}
\end{table*}
\subsection{Total Mass Distribution}
\label{global-mass}
Abell 2744 is a massive and highly dynamically disturbed galaxy cluster \citep[][]{giovannini99,govoni01a,kempner04,zhang04,owers11,girardi01,boschin06,braglia07}. 
Following the method presented in Sect.~\ref{slwl_model}, we reconstructed its mass distribution within a field of view of $\sim 4\,$Mpc$^{2}$ ($\sim$2~Mpc radius). Fig.~\ref{a2744_colour} shows the inner $\sim$1~Mpc$^{2}$, the region where susbtructures are detected with relatively high significance. Mass contours are drawn in white on Fig.~\ref{a2744_colour}.
We obtain a highly clumpy mass distribution in this region, with numerous substructures present within 1~$h^{-1}$Mpc from the cluster centre -- defined here as the position of the BCG ($\alpha$=3.586259 $\delta$=-30.400174). We measure a total mass $M(R<1000\, {\rm kpc}) = 1.85\pm 0.07\times 10^{15}\, M_{\odot}$.

\cite{merten11} (M11 hereafter) present the most detailed gravitational lensing analysis of Abell 2744 before the HFF and deep \emph{XMM-Newton} data were taken, measuring a total mass of the cluster of $M_{M11}(R < 1.3\, {\rm Mpc}) = 1.8\pm 0.4 \times 10^{15}\, M_{\odot}$.
More recently, \cite{medezinski16} (M16 hereafter) measured a total mass of $M_{M16}(R<1.3\, {\rm Mpc}) = 1.65\pm 0.23\times 10^{15}\, M_{\odot}$.
Both values are lower than our estimate, $M(R<1.3\, {\rm Mpc}) = 2.3\pm 0.1 \times 10^{15}\, M_{\odot}$, agreeing within the error with M11 but not with M16. This latter disagreement may be due to the lack of strong-lensing constraints in the M16 mass modeling, thus leading to an under-estimate of the total mass of the cluster.

\subsection{Substructure Mass Distribution}
\label{substr_mass}
With the strong+weak lensing analysis presented here, we detected 8 substructures within 1\,Mpc from the cluster BCG, including the main cluster halo. Their coordinates, masses (within a 150~kpc aperture), mass-to-light ratio (M/L afterwards), and distance to the cluster centre are listed in Table~\ref{tab_substr}, and highlighted with orange diamonds on Fig.~\ref{a2744_colour}.
To measure the M/L ratio of the substructures, we used the method presented in \cite{jauzac15a} for MACSJ0416, and looked for galaxies within $R<150$kpc from the mass peak. All substructures detected by M11 and/or M16 are discussed below, and masses are quoted within an aperture of 250~kpc for comparison with these two analyses (see also Table~\ref{tab_substr2}).
In Table~\ref{tab_substr} masses are quoted within a smaller aperture (150~kpc) than in Table~\ref{tab_substr2}, as some of the substructures are quite close to each other, and thus using a larger aperture would lead to an over-estimate of the mass, i.e. taking into account some of the mass from neighboring substructures. Thus \textit{`this work'} values quoted in Table~\ref{tab_substr2} should be taken with caution.

M11 presented the first strong+weak lensing analysis of Abell 2744, while discussing the detection of multiple substructures around the core of the cluster, and was followed more recently by a weak-lensing only analysis by M16.
M11 present a multiple merger, with four cluster-scale components within $\sim$700~Mpc to the cluster centre: \emph{Core}, \emph{North} (N), \emph{North-West} (NW) and \emph{West} (W).
While the \emph{Core} was imaged with HFF, the N and NW components are visible on the pre-HFF ACS images. Both reveal strongly-lensed objects around their BCGs confirming the presence of two relatively massive substructures. However, the lack of spectroscopy for these lensed objects makes their strong-lensing mass modeling difficult due to degeneracies. Therefore, M11 only studied them using weak-lensing. The W substructure was discovered by their \emph{Subaru} weak-lensing analysis as it lies outside the HST coverage. 
The measured masses within an aperture of $R=250$~kpc are listed in Table~\ref{tab_substr2}.

M16 present the weak-lensing analysis of Abell 2744 using more recent and deeper Subaru data compared to M11.
They detect 4 substructures with signal-to-noise ratio (S/N) greater than 4.5 named, \emph{Core}, \emph{W}, \emph{NE}, and \emph{NW}. Both the \emph{Core} and \emph{NW} components correspond to the one detected in M11. However, the \emph{W} halo is detected much closer to the centre of the cluster ($\alpha_{W, M16} = 3.556083$, $\delta_{W, M16} = -30.399277$), which is not consistent with the position given in M11 ($\alpha_{W, M11} = 3.5291667$, $\delta_{W, M11} = -30.406667$). The NE substructure is a new detection. Masses are also given in Table~\ref{tab_substr2}.

The \emph{Core} component of the cluster is modeled using the HFF strong-lensing constraints. We measure a mass of $M_{Core} (R<250\, {\rm kpc}) = 2.77\pm 0.01\, \times 10^{14}\, M_{\odot}$, in agreement with measurements in M11, however much larger than the M16 estimate of $M_{Core, M16} (R<250\, {\rm kpc}) = 1.49\pm 0.35\, \times 10^{14}\, M_{\odot}$. We attribute this difference to the fact that M16 do not include strong-lensing in their modeling, therefore their estimate of the mass within 250\, kpc from the \emph{Core} BCG (highly non-linear region) is an extrapolation from their weak-lensing measurement which may lead to an under-estimation of the mass enclosed in this region. 

The \emph{North} substructure is detected at the same position as M11, however our mass model reveals a much more massive substructure than M11, $M_{N} (R<250\, {\rm kpc}) = 1.47\pm 0.09\, \times \, 10^{14}\, M_{\odot}$. The \emph{N} component is not detected by the weak-lensing analysis in M16.

M11 \emph{NW} substructure is in fact composed of two mass peaks, named \emph{NW1} and \emph{NW2} in their paper.
M16 also detect a \emph{NW} substructure, elongated along the same direction and that seems to be composed of two unresolved halos.
Our mass reconstruction reveals two substructures at the position of \emph{NW1} and \emph{NW2} halos, called here \emph{NW} and \emph{S3} respectively. Moreover, the mass map reveals a third halo, \emph{S4}, aligned with \emph{NW} and \emph{S3}. These three substructures extend in the direction of the NW filament detected in \cite{eckert15} (see Fig.~\ref{a2744_colour}).
The \cite{owers11} spectroscopic redshift catalogue allowed us to identify 10, 3 and 2 objects as being at the same location as the \emph{NW}, \emph{S3}, and \emph{S4} halos respectively (within $150\, {\rm kpc}$ of their mass peak). Despite the low statistics, these three substructures seem to be at the cluster redshift, as are all spectroscopically identified galaxies.
Another point that will be discussed in more detail in Sect.~\ref{dynamics} is the position of \emph{S4}. Indeed, \emph{S4} coincides with a remnant-core detected in the \emph{Chandra} data (X2 in Fig.~\ref{a2744_colour}). In M11, no dark matter counterpart was found for this X-ray peak, and this structure was then 'nick-named' the \emph{interloper}. Our analysis contradicts M11's interpretation. 

While the \emph{Core}, \emph{N}, and \emph{NW} substructures from M11 are recovered by our mass reconstruction, the \emph{W} substructure, nick-named the \emph{Ghost clump} by M11 because of a lack of X-ray counterpart, is not detected in our analysis.
Fig.~\ref{a2744_colour} shows the position of M11's \emph{W} substructure as a red diamond, and as one can see while looking at the mass contours, no clear mass peak is detected at this location. The enclosed mass we measure at the M11's position is $M_{W, Jauzac+16} (R<250\,{\rm kpc}) = 0.39\pm 0.08 \times 10^{14}\, M_{\odot}$, almost a factor of 4 lower than M11's estimate. 
This \emph{W} substructure location coincides with a bright cluster galaxy, that was hypothesized as being the BCG of the structure in M11. The galaxy is included in our model, but the shear signal around it does not indicate any substructure as massive as the one detected by M11. The lack of a corresponding dark-matter halo at the position of the M11 \emph{W} structure could indicate that the associated galaxies have already merged with the main cluster halo, and thus that their dark-matter counterpart has been stripped.
As M11 \emph{W} is not detected by M16 either, we can conclude that this substructure is an artefact created by the mass reconstruction and the shallow Subaru data used in the M11 analysis. However, we detect a smaller substructure closer to the cluster \emph{Core}, named $W_{bis}$ (see Table~\ref{tab_substr} and Fig.~\ref{a2744_colour}), at a distance of $D_{W_{bis}-W} = 247\,{\rm kpc}$ from the M11 \emph{W} substructure. 
Its position is consistent with the M16 \emph{W} substructure within errors.
To quantify $W_{bis}$, we looked for corresponding spectroscopic counterparts from the \cite{owers11} spectroscopic redshift catalogue. Within a radius of $R=150\, {\rm kpc}$ from its mass peak, we find only 2 galaxies, both being background objects (identified following \cite{owers11} cluster membership criteria presented in Sect.~\ref{zspec_obs}).
Using the WFI cluster member catalogue presented in Sect.~\ref{CM_model}, we estimate the M/L$_{K}$ for the $W_{bis}$ substructure of $\sim 700$, in agreement with the M16 estimation of $584\pm162$.
While the statistic does not allow us to firmly conclude anything about $W_{bis}$, its M/L and spectroscopic redshift indicate that it is a background structure, still detected in our lensing mass reconstruction as it provides us with a 2D mass reconstruction encompassing all structures along the line of sight.





The \emph{S1} substructure, located North-East of the cluster \emph{Core}, corresponds to the \emph{NE} substructure detected by M16 (see red diamond on Fig.~\ref{a2744_colour}). It aligns with the gas bridge found by \cite{eckert16} that relates the radio relic to the cluster \emph{Core}. The matching with the \cite{owers11} spectroscopic catalogue reveals 4 galaxies, all at the redshift of the cluster, and 26 galaxies identified as cluster galaxies with the WFI data. 
Finally, the \emph{S2} substructure corresponds to a clear light peak, with 29 galaxies in the WFI cluster member catalogue, as well as 2 galaxies identified in the \cite{owers11} catalogue. As for \emph{S1}, \emph{S3} and \emph{S4}, despite the low number of spectroscopic counterparts, all of them seem to be at the cluster redshift. 


\begin{table}
\begin{center}
\begin{tabular}[h!]{cccc}
\hline
\hline
\noalign{\smallskip}
$ID$ & $M_{this\, work}$ & $M_{Merten+11}$ & $M_{Medezinski+16}$ \\
\hline
\emph{Core} & $27.7 \pm 0.1$ & $22.4\pm 5.5$ & $14.9\pm 3.5$ \\
\emph{N} & $14.7 \pm 0.9$ & $8.6\pm 2.2$ & --  \\
\emph{NW} & $18.0 \pm 1.0$ & $11.5\pm 2.3$ & $7.6\pm 3.5$ \\
$W_{bis}$ & $12.9 \pm 1.1$ & $11.1\pm 2.8$ $^{\ast}$ & $12.5\pm 3.5$ \\
S1 & $13.0 \pm 1.0$ & -- & $9.5\pm 3.5$ $^{\ast \ast}$ \\
\noalign{\smallskip}
\hline
\hline
\end{tabular}
\caption{Masses of substructures detected in this work, \citet{merten11} and \citet{medezinski16}, in units of $10^{13}$M$_{\odot}$. Masses are quoted in a 250~kpc aperture for comparison with \citet{merten11} and \citet{medezinski16} values. $^{\ast}$ The mass given here corresponds to the \emph{W} substructure of \citet{merten11}.
The $W_{bis}$ substructure is consistent with the \citet{medezinski16} \emph{W} substructure. $^{\ast \ast}$ The \emph{S1} substructure is consistent with the \citet{medezinski16} \emph{NE} substructure.
}
\label{tab_substr2}
\end{center}
\end{table}

\section{X-ray Analysis}
\label{xray_thermo}

\begin{figure}
\includegraphics[width=0.5\textwidth]{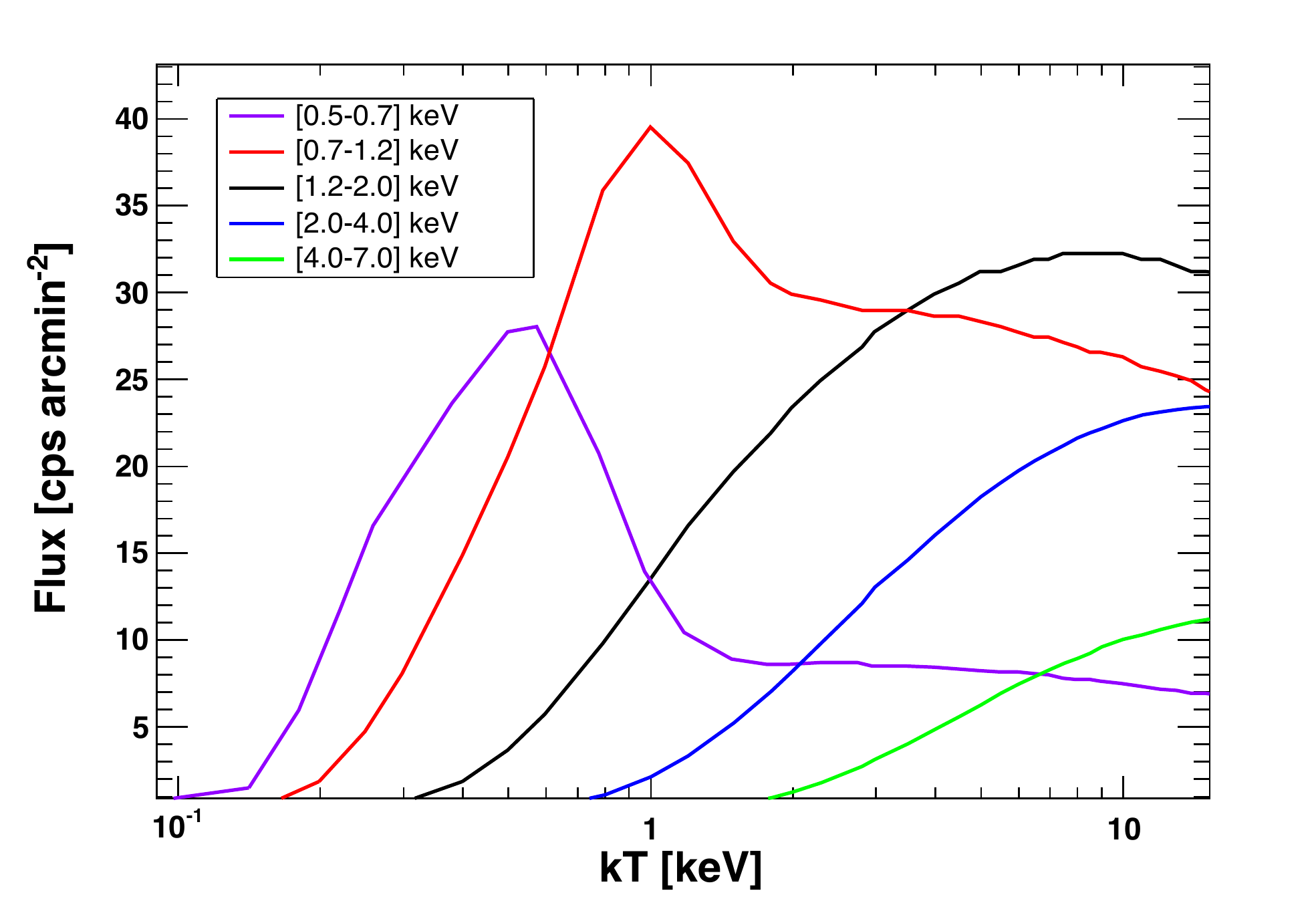}
\caption{
Spectral templates used for the production of the thermodynamic maps shown in Fig. \ref{fig:thermo}. The various curves show the model \emph{XMM-Newton}/EPIC count rate in each of our five energy bands (see legend) as a function of temperature.
}
\label{fig:xraytemplates}
\end{figure}

To study the state of the hot ICM of A2744, we used our \emph{XMM-Newton} data to extract thermodynamic maps of the central regions of the cluster. To this aim, we use the \emph{XMM-Newton} images in 5 energy bands spanning the [0.5-7] keV range (see Sect. 2.2) and the corresponding exposure maps and background maps. The intensity of the sky background is estimated in each band by computing the mean surface brightness in a source-free region. Assuming that the X-ray emission can be locally described by a single-temperature absorbed APEC model \citep{smith01xray}, we use XSPEC to fold the model with the \emph{XMM-Newton} response. We fix the metallicity to the canonical value of $0.3\, Z_\odot$ \citep{leccardi08} and the Galactic column density to the value of $1.5\times10^{20}$~cm$^2$ estimated from 21cm maps in the direction of A2744 \citep{kalberla05}. We then vary the plasma temperature to predict the expected count rate in each of our five bands as a function of temperature and create templates relating count rate to temperature. In Fig. \ref{fig:xraytemplates} we show the expected count rate at fixed emission measured in the five energy bands.

\begin{figure*}
\includegraphics[width=0.5\textwidth]{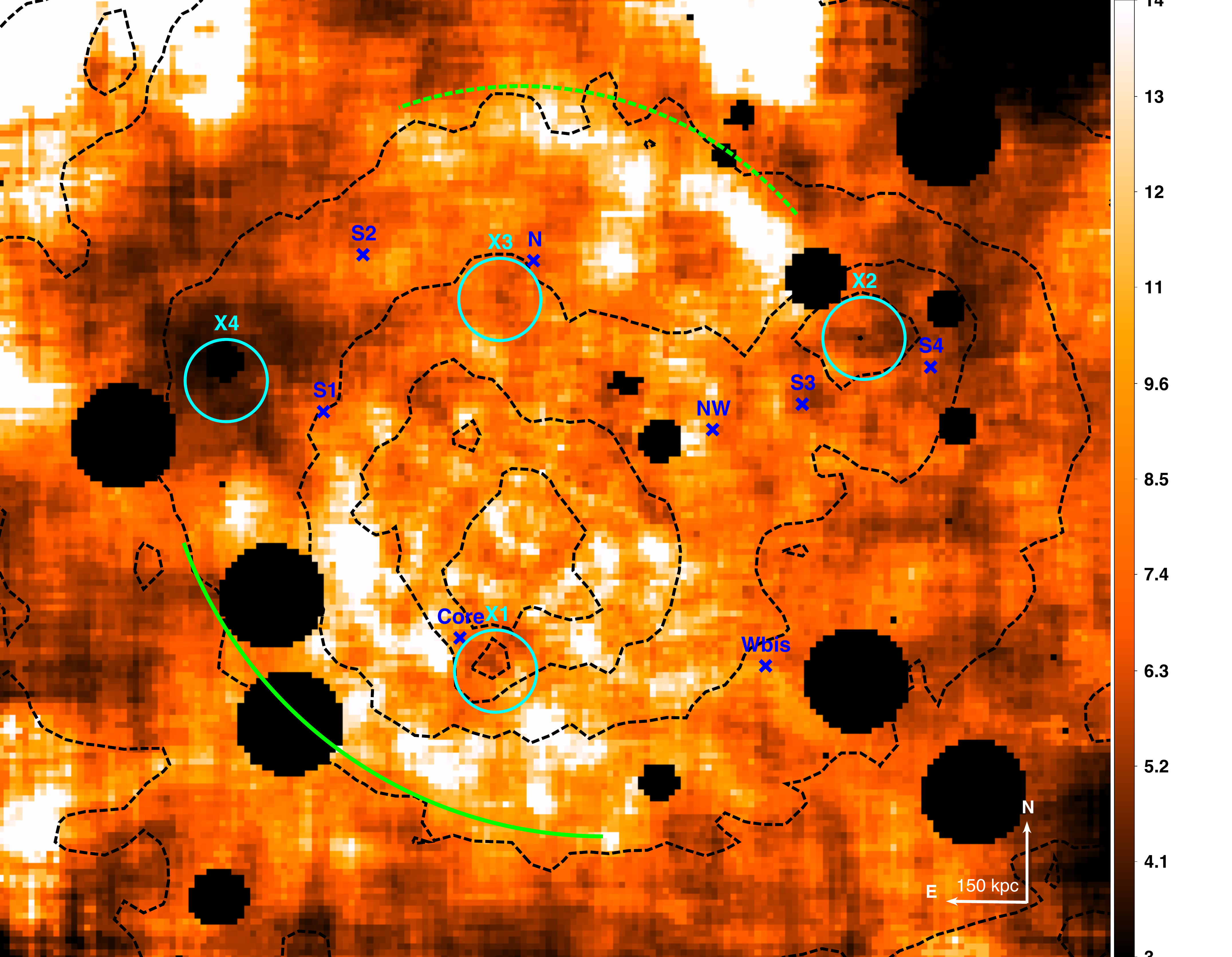}\includegraphics[width=0.5\textwidth]{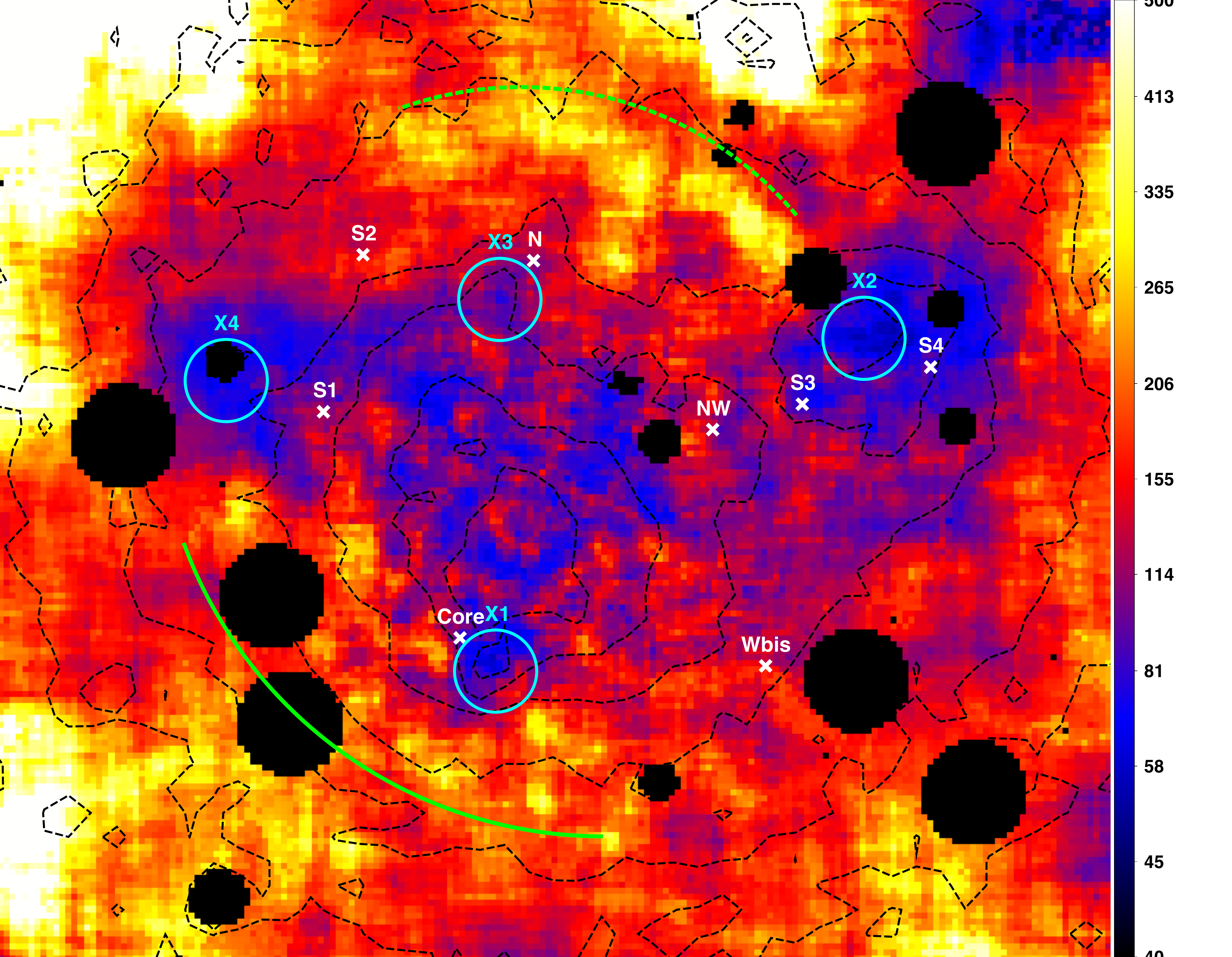}
\caption{
\emph{XMM-Newton} maps of temperature (left) and pseudo-entropy (right) for A2744. The dashed black contours denote X-ray surface brightness in the [0.7-7] keV range as observed with \emph{Chandra}. The cyan circles show the position of low-entropy cores identified in our analysis; the crosses indicate the position of the substructures listed in Table~\ref{tab_substr}. The arc-like feature located South-East of the cluster core and denoted by the solid green line shows the shock front identified by \citet{owers11} in \emph{Chandra} data, whereas the dashed green line North of the cluster core shows a second putative front moving in the opposite direction (see text).
}
\label{fig:thermo}
\end{figure*}

Around each pixel, we define a circular region containing at least 500 counts in the full [0.5-7] keV band, and we measure the vignetting-corrected and background-subtracted count rate in each of our five bands. We then use a maximum-likelihood algorithm to fit the model APEC templates shown in Fig. \ref{fig:xraytemplates} to the five data points and provide an estimate of the local temperature and emission measure with their uncertainty. We then construct a temperature map by gathering the best-fit values for each pixel. We also compute a (projected) entropy map by combining the local best-fit temperature and emission-measure values and computing the (pseudo-)entropy $K=kT\times(EM)^{-1/3.}$. The resulting thermodynamic maps are shown in Fig. \ref{fig:thermo}. Note that the values of neighboring pixels obtained through this technique is obviously correlated, with a correlation length given by the local data quality; the correlation length ranges from a size similar to the \emph{XMM-Newton} PSF ($8^{\prime\prime}\sim30$ kpc) in the central regions to $\sim40^{\prime\prime}$ (160 kpc) in the outermost regions shown in Fig. \ref{fig:thermo}. 

\section{Dynamics}
\label{dynamics}

\begin{table}
\caption{\label{tab:remnant}Position of the remnant low-entropy cores identified in our X-ray analysis.}
\begin{center}
\begin{tabular}{ccc}
\hline
\hline
ID & R.A. (deg) & Dec. (deg)\\
\hline
X1 & 3.5826068 & -30.402491 \\
X2 & 3.5339304 & -30.366334\\
X3 & 3.580766 & -30.360742\\
X4 & 3.6168333 & -30.371133\\
\hline
\hline
\end{tabular}
\end{center}
\end{table}

In Fig. \ref{fig:thermo} we highlight the position of several features observed in our thermodynamic maps. In particular, the position of four individual low-entropy cores is shown by the cyan circles. 

\subsection{The remnant cores}

The X-ray peak associated with the main core (X1) was discussed in detail by \citet{owers11} and M11; the X-ray peak is offset by 25 arcsec (120 kpc) from the main mass peak. The redshift distribution in this region was found to be bi-modal, with the high-velocity component (2,500 km/s) interpreted as a bullet-like component observed close to the line of sight. 

The prominent X2 feature located North-West of the cluster core \citep[dubbed the \emph{interloper} by][]{owers11} has a mean gas temperature of $\sim5$ keV, in agreement with previous studies \citep{kempner04,owers11}. While the temperature and size of this gas structure associate it unequivocally with a massive subcluster with a mass in excess of $10^{14}M_\odot$, M11 found that the main associated mass peak (consistent with our NW clump) is located more than 300 kpc in projection from the peak of the gas structure. However, our analysis reveals the high-confidence detection of an additional cluster-size halo (S4) consistent with the position of the X-ray peak (see Fig. \ref{a2744_colour}), which contradicts this interpretation. This substructure was marginally detected as an overdensity in the photometric catalogue of \citet{owers11}. The presence of a trail of cool gas located South of this substructure and of a cold front to the North indicate that this substructure is in an early stage of merging and is currently moving towards the North direction.

The third structure (labelled X3 in Fig. \ref{fig:thermo}) was identified by \citet{owers11} and is detected as well in our analysis. This structure is located 18 arcsec (80 kpc) from the North clump (see Fig. \ref{a2744_colour}) and it is followed by a plume of low-entropy gas extending South of the mass peak \citep{owers11}, which indicates that it is moving in the North direction. The relatively low surface brightness of the X-ray structure and the offset between gas and DM suggest that this clump is in an advanced stage of merging and that most of the associated gas has been stripped from its original halo.

Finally, we report the high-confidence detection of an additional, previously unreported structure (labelled X4 in Fig. \ref{fig:thermo}) located 2.5$^\prime$ (700 kpc) North-West of the cluster core. The gas of this structure has a mean temperature of $\sim3.5$ keV and its entropy is the lowest of the cluster. This low-entropy core is located 40 arcsec East of the massive substructure S1 (see Fig. \ref{a2744_colour}). 

The substructures labelled as NW, W, S2, and S3 do not have any obvious X-ray counterpart in our thermodynamic maps. This suggests that these structures are the remnants of previous merging activity and that the gas originally present within these massive clumps has been completely stripped and virialized.

\subsection{Shock fronts and dynamics}

A2744 is known to host several well-documented shock fronts induced by its dynamical activity. The most prominent feature located SE of the core (see Fig. \ref{fig:thermo}) was originally reported by \citet{owers11} and it is associated with a density jump $n_{\rm in}/n_{\rm out}\sim1.6$, corresponding to a Mach number of $\sim1.4$. Our analysis clearly highlights the presence of the shock-heated gas in the core region. In case this feature is caused by the motion of the main core, this suggests that the core is currently moving in the SE direction, as originally noted by \citet{owers11}. 

Recently, a second shock front located 1.5 Mpc NE of the cluster core was reported by \citet{eckert16}. The shock front is associated with the Eastern edge of the radio relic \citep{orru07}. While this feature is located outside of the region studied here, its location (see Fig. \ref{a2744_colour}) may suggest that it is associated with the motion of substructure S1. In this case, substructure S1 would be moving toward the NE direction after a first core passage.

Additionally, our thermodynamic maps indicate the presence of another high-entropy arc-like feature located $\sim300$ kpc north of the N substructure (see Fig. \ref{fig:thermo}). The temperature beyond the arc falls sharply from $\sim14$ keV to $\sim6$ keV. While these properties are suggestive of an additional shock front, no brightness edge is observed in the high-resolution \emph{Chandra} map at this position. In case this feature is a true shock front, the absence of a coincident brightness jump indicates that the front would be traveling at an angle which is highly inclined with respect to the plane of the sky, washing out the brightness edge. If this interpretation is correct, the association with the N structure and the location of this front opposite to the SE shock would reinforce the interpretation of \citet{owers11} that the main merger direction is occurring along the N-S axis.

\section{Discussion}
\label{discussion}

\subsection{Comparison with the Millennium XXL simulation}

\begin{figure}
\includegraphics[width=\columnwidth]{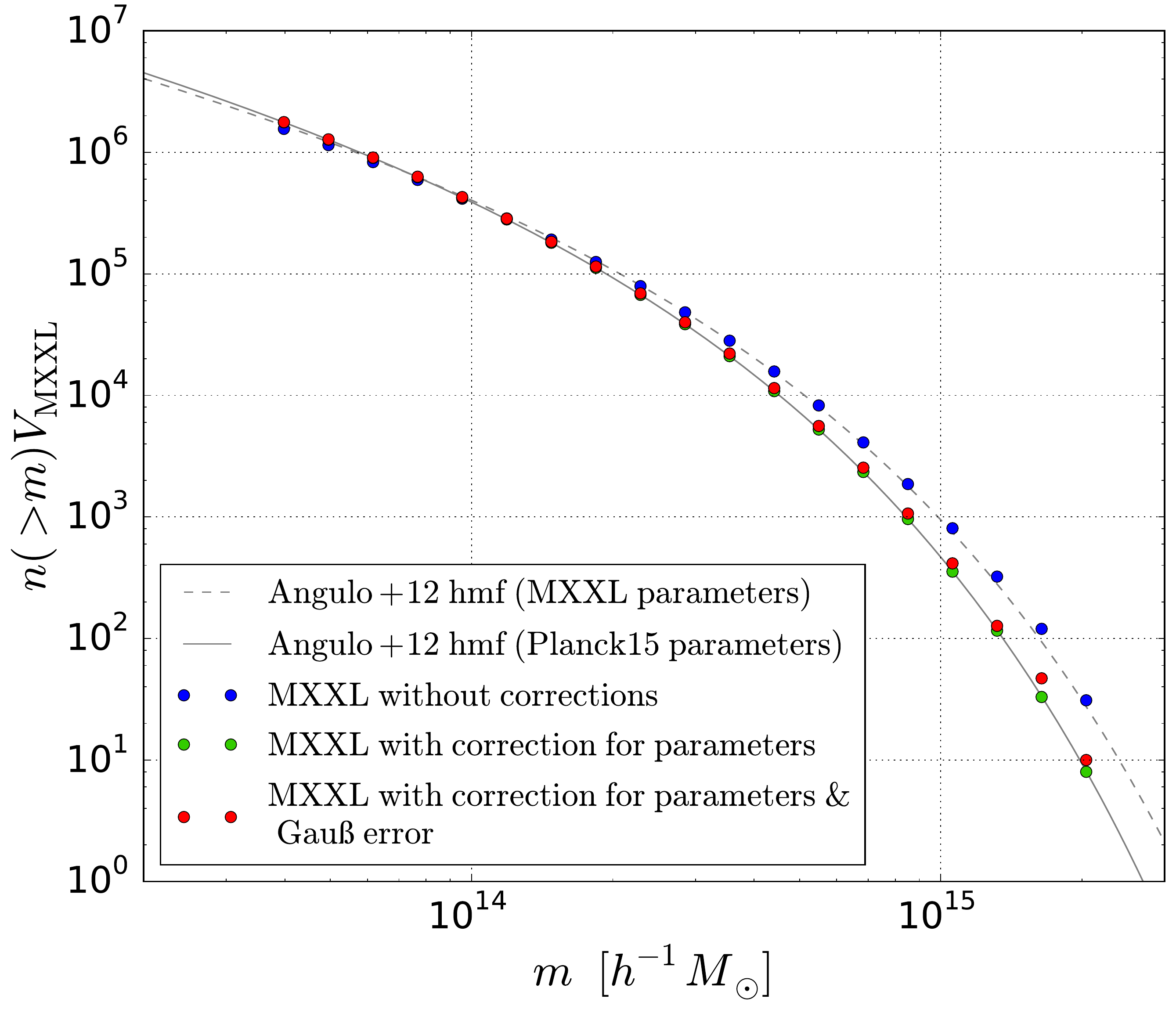}
\caption{
Cumulative subhalo mass distribution. The lines show the theoretical halo mass functions for subhaloes with the original MXXL parameters (\emph{dashed}) and the Planck 2015 parameters (\emph{solid}). The dots show the cumulative mass function calculated from MXXL subhaloes without corrections (\emph{blue}), after correcting the mass distribution to reflect updated cosmological parameters (\emph{green}), and after additionally correcting for mass uncertainties in the lensing analysis (\emph{red}).
}
\label{hmf}
\end{figure}

\subsubsection{The Millennium XXL simulation}
We compare the observations discussed above with theoretical predictions from the Millennium XXL (MXXL) simulation \citep[][]{Angulo2012}. This simulation models structure formation in a $\Lambda$CDM Universe within a cube of size ($3\,h^{-1}{\rm Gpc}$)$^3$ using particles of mass $m_{p} = 6.16 \times 10^{9}\, h^{-1}{\rm M}_{\odot}$ and cosmological  parameters of $H_0~=~73\,{\rm km\,s^{-1}Mpc^{-1}}$, $\Omega_\Lambda = 0.75$, $\Omega_{\rm m} = \Omega_{\rm dm} + \Omega_{\rm b} = 0.25$, $\Omega_{\rm b} = 0.045$ and  $\sigma_8 = 0.9$, matching those used in the original Millennium simulation \citep[][]{springel05}. 

Within MXXL, gravitationally bound structures are identified at two hierarchical levels: dark-matter haloes, found using a Friends-of-Friends (FoF) algorithm \citep[][]{Davis1985}, and gravitationally bound substructures within these FoF haloes, identified using the {\tt SUBFIND} algorithm \citep[][]{Springel2001}. Since, out to a redshift of $z{=}0.308$, the volume of the MXXL simulation exceeds that of the real Universe by over an order of magnitude, the odds are enhanced of being able to find a small number even of very rare objects. We perform our search for a structure mimicking Abell 2744 within the MXXL snapshots at redshift $z{\approx}0.28$ and $z{\approx}0.32$, bracketing the redshift of Abell 2744 of $z{=}0.308$.

\subsubsection{Searching for an MXXL version of Abell 2744}
We use two different cluster properties to quantify the probability of a cluster like Abell 2744 existing in a $\Lambda$CDM Universe: \emph{(1)} its total mass, and \emph{(2)} the substructures detected in its surroundings, i.e., the masses of subhaloes and distances between them.

Before comparing the observed properties of Abell 2744 with those of MXXL clusters we first adjust the masses of the MXXL haloes to match the halo mass function (see Fig.\,\ref{hmf}) for the latest cosmological parameters as measured by \emph{Planck} \citep[][]{Planck2015}.
This step is necessary as the MXXL simulation uses the same cosmological parameters as the previous two Millennium runs \citep[][]{Springel2005,Boylan-Kolchin2009}.
We perform this adjustment by sorting all FoF haloes by mass down to $10^{13}\, h^{-1}{\rm M}_{\odot}$ and then assigning each FoF halo the mass at the according rank in the theoretical cumulative mass distribution. We use the form of the halo mass function obtained by a fit to the subhaloes in all three Millennium runs \citep[Eq.~3,][]{Angulo2012}. 
A similar adjustment is made for the mass of subhaloes identified by {\tt SUBFIND}. Either halo mass function is calculated at redshift $z=0.28$ and $z=0.32$ with \emph{Planck  2015} parameters using the python module {\tt hmf} \citep[][]{Murray2013}, which contains implementations of both \citet{Angulo2012} mass functions.

Fig.\,\ref{hmf} shows that the changes in $\Omega_m$ and $\sigma_8$ move the halo mass function in different directions. The smaller value of $\sigma_8$ from \emph{Planck} results in a lower abundance of massive haloes, while the increased value of $\Omega_m$ causes an overall shift of the mass function to higher values, which visibly affects masses below $10^{14}\,h^{-1}{\rm M}_{\odot}$. By design, the adjusted halo masses follow the updated version of the halo mass function (green data points in Fig.\,\ref{hmf}).

Next, we try to account for errors in the masses measured in the gravitational-lensing analysis by conservatively adopting the largest uncertainty found in our analysis (see Sect.~\ref{lensing_mass}) of approximately 15\% as a universal relative mass error. We then draw for each subhalo a new mass from a Gaussian distribution with mean of $m_{\textrm{corr}}$ (given by the correction with respect to updated cosmological parameters) and standard deviation given by the relative mass error of 15\%. Accounting for the resulting scatter has a noticeable effect on the mass function. This well established effect, known as Eddington bias \citep[][]{Eddington1913}, stems from the fact that the slope of the halo mass function steepens with increasing mass. As a result, erroneous up-scattering of intrinsically low-mass haloes will outweigh down-scattering of intrinsically high-mass haloes, causing an appreciable increase in the number of high-mass haloes. By contrast, Eddington bias is barely noticeable at masses $<4\times 10^{14}\,h^{-1}{\rm M}_{\odot}$ where the slope of the halo mass function is nearly constant.

Having applied the halo-mass corrections described above, we search for FoF-haloes with masses within $3\sigma$ of the total mass of Abell 2744 ($M(R<1.3{\rm Mpc}) = 2.3\pm0.1\times 10^{15}\, {\rm M}_\odot$) as determined in Sect.~\ref{lensing_mass}. Since FoF haloes in the MXXL simulation can be considerably larger in size than our measurement aperture of $1.3{\rm Mpc}$, we use an NFW-profile \citep[]{NFW96} to extrapolate the measured mass to the size of an MXXL FoF halo. To ensure that we are comparing like and like, we integrate the NFW profile over a cylinder with radius $r = 1.3\,{\rm Mpc}$ and length\footnote{The cylinder length of $l = 30\,{\rm Mpc}$ was chosen because $\rho_{\rm NFW}$ drops at a radius of $r \approx 15\,{\rm Mpc}$ below the mean matter density.} $l = 30\,{\rm Mpc}$ thus including any line-of-sight projection in mass that we also would have included in a gravitational-lensing analysis. 
Using the concentration-mass relation presented in \citet{neto07} (Eq.\ 4), we calculate $M_{200}$ (i.e., the mass within a sphere with a density equal to or higher than 200 times the critical density of the Universe) such that the mass in the cylinder matches the measured mass of Abell 2744 of $M(R<1.3{\rm Mpc}) = 2.3\pm0.1\times 10^{15}\, {\rm M}_\odot$. 
Finding the mode of the distribution of MXXL FoF masses to be roughly 22\% higher than $M_{200}$ \citep[for a similar estimation see ][]{jiang14}, we thus increase the $M_{200}$ values by 22\% to obtain the final mass to search for in the MXXL simulations: $M_{\rm FoF} = 4.0\pm0.2\times 10^{15}\, {\rm M}_\odot$. We find $39$ MXXL clusters at either $z\approx0.32$ or $z\approx0.28$ and conclude that clusters with a mass comparable to that of Abell 2744 are common in the MXXL simulation. 

To assess the second property of Abell 2744, the number and mass of substructures, we analyse the properties of MXXL subhaloes identified by {\tt SUBFIND}. Analogously to the method described above, we extrapolate the {\tt SUBFIND}-mass of subhaloes in MXXL by integrating a NFW-profile over a cylinder with radius $r = 150\,{\rm kpc}$ and length $l = 30\,{\rm Mpc}$. 
Like for the FoF haloes, we also assume the {\tt SUBFIND} mass to be 22\% higher than $M_{200}$. 
Table~\ref{substr_nfw} gives the NFW-extrapolated masses of the 8 substructures discussed in Sect.~\ref{substr_mass}.
Within a box centred on each subhalo, we then count the number of subhaloes with masses comparable to extrapolated subhalo masses. Since gravitational lensing measures the projected mass, we choose the box to be considerably deeper ($15\,h^{-1}{\rm cMpc}$) than wide ($2\,h^{-1}{\rm cMpc}$) and only consider the projected 2D distances. The value of $15\,h^{-1}{\rm cMpc}$ was chosen because it is representative of redshift-space distortions. 
The search is performed for three different orientation, i.e., adopting each of $x$, $y$, and $z$ as the line of sight. Since this process is computationally expensive, we only consider subhaloes with masses above $3.5\times 10^{13}\, h^{-1}{\rm M}_{\odot}$.

We find no cluster in the MXXL simulation with a substructure distribution similar to that of Abell 2744 (eight extrapolated subhalo masses above $10^{14}\, {\rm M}_{\odot}$, all within a radius of $1\,{\rm Mpc}$ around the centre of mass). None of the 39 MXXL clusters identified as featuring a total mass similar to that of Abell 2744 have more than three subhaloes with a mass above $10^{13}\, {\rm M}_{\odot}$ within a radius of $1\,{\rm Mpc}$ around the centre of mass. Instead, all of these MXXL clusters have a massive core with $M_{\rm Core} \approx 3 - 4 \times 10^{15}\, {\rm M}_{\odot}$ and in most of the cases around ten subhaloes with masses of $10^{11} - 10^{13}\, {\rm M}_{\odot}$ within a radius of 1~Mpc. Increasing the depth along which the clusters are projected to an implausible value of $30\,h^{-1}{\rm cMpc}$ does not turn up any Abell 2744 contenders either. Indeed, it is improbable that subhaloes have met within the lifetime of the Universe when distributed over such large distance.

\subsubsection{Is Abell 2744 compatible with $\Lambda$CDM?}

As discussed in the preceding section, we find that clusters as massive as Abell 2744 are common in the MXXL simulation, in agreement with other studies investigating the compatibility of very massive clusters with $\Lambda$CDM \citep[e.g.][]{Hotchkiss2011,Waizmann2013}. Abell 2744 might pose a challenge to $\Lambda$CDM nonetheless though, as we fail to find massive clusters with a similarly high number of massive subhaloes at a close distance from the halo centre in the MXXL simulations.

Nonetheless, our comparison with MXXL contains some caveats. The first concerns the old set of cosmological parameters used in the MXXL simulation. This affects the outcome in two ways: (1) the halo mass function changes and therefore halo masses themselves are different; (2) merging scenarios and their time scales are influenced. We tried to take the former into account by modifying the masses such that the halo mass function fits that obtained with the \emph{Planck} cosmological parameters. On the other hand, the impact on the merging scenarios is not considered here, but could still affect the outcome considerably. This becomes obvious when looking at the merger rate presented in \citet[][]{Lacey1993}.
Adapting their discussion, the instantaneous merger probability, i.e. the probability that a halo of mass $M_1$ merges with a halo of mass $\Delta M$ into a halo of mass $M_2 = M_1 + \Delta M$ within a scale factor change of $\mathrm{d}\ln a$, is given by 
\begin{equation}
\begin{aligned}
\frac{\mathrm{d}^2 p}{\mathrm{d}\!\ln\Delta M\,\mathrm{d}\!\ln a} =  &\left(\frac{2}{\pi}\right)^{1/2}\left|\frac{\mathrm{d} \ln \delta_c}{\mathrm{d} \ln a}\right| \left(\frac{\Delta M}{M_2}\right) \\
& \times \left|\frac{\mathrm{d} \ln \sigma_2}{\mathrm{d} \ln M_2}\right|\frac{\delta_c(a)}{\sigma_2} \frac{1}{(1-\sigma_2^2/\sigma_1^2)^{3/2}} \\
& \times\exp\left[-\frac{\delta_c(a)^2}{2}\left(\frac{1}{\sigma_2^2}-\frac{1}{\sigma_1^2}\right)\right]\ ,
\end{aligned}
\end{equation}
where $\sigma_1 \equiv \sigma(M_1)$ and $\sigma_2 \equiv \sigma(M_2)$ are the density contrast variances after a smoothing with a window function containing mass $M_1$ or $M_2$, respectively, and $\delta_c(a)$ denotes the critical density contrast at scale factor $a$ at which a region collapses according to spherical top-hat collapse.
The instantaneous merger probability for two haloes of masses $M_1 = \Delta M = 10^{14}\,h^{-1}{\rm M}_{\odot}$ at redshift $z = 0.308$ increases by 3\% from a value of $0.456$ with the MXXL parameters to a value of $0.470$ with the \emph{Planck} parameters. The rate of merger events at the investigated time scale and mass range is therefore underpredicted in MXXL in comparison to a universe with \emph{Planck} parameters. However, it is important to note that the merging probability is not increased at all time scales. In fact, the integrated merging probability $\mathrm{d}p/\mathrm{d}\!\ln \Delta M$ decreases by 11\% from $0.456$ to $0.408$ with up-to-date parameters.

The second difficulty is the comparison of gravitational lensing masses with halo masses derived from the MXXL simulation. We extrapolate the FoF-mass and \texttt{SUBFIND}-masses of the main halo and substructures using NFW profiles, and consider errors on lensing mass measurements. Line of sight projection of several clusters within the aperture also leads to an add-up scattering in mass.
Allowing the masses within the aperture to add up in the MXXL analysis, we find that on average 1.3 subhaloes per FoF-halo are scattered to masses above $10^{14} {\rm M}_\odot$. This shows that projection effects within the aperture do have a noticeable impact but these alone cannot explain all of the massive substructures.

\begin{figure*}
\includegraphics[width=0.875\textwidth]{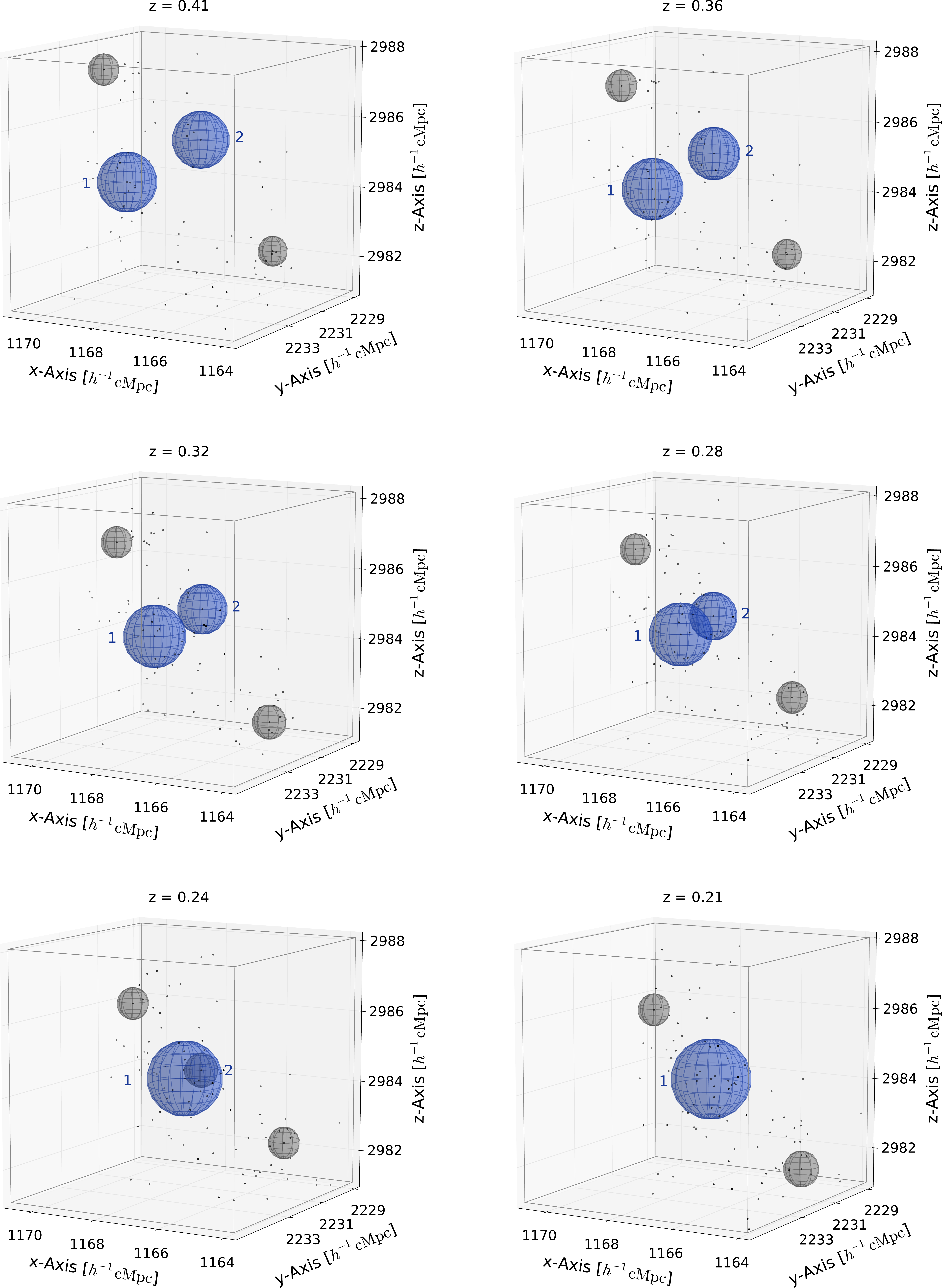}
\caption{
A schematic of the galaxy cluster in the MXXL most like Abell 2744 (but still containing much less substructure). 
Two subhaloes of the simulated cluster merge at redshift $z=0.41$, having started with masses $M_1 = 0.857\times 10^{14}\, h^{-1}{\rm M}_{\odot}$ and $M_2 = 0.857\times 10^{14}\, h^{-1}{\rm M}_{\odot}$. The radius of the spheres corresponds to the $r_{200}$ value estimated from the mass of the halo according to Eq.~\ref{eq:r200}. Subhaloes with masses $<10^{13}\, h^{-1}{\rm M}_{\odot}$ were plotted as dots for clarity. The masses of both subhaloes for each snapshot are listed in Table~\ref{mergerMasses}.
}
\label{merger}
\end{figure*}

Thirdly, it should be mentioned when looking at the substructure distribution, that numerical effects on subhalo detection can have an important impact. When monitoring the time evolution of a merger between two subhaloes (both having a mass of about $10^{14}\, h^{-1}{\rm M}_{\odot}$), it turns out that the mass of the smaller halo decreases rapidly once it reaches a distance of $1.5-2\ h^{-1}{\rm Mpc}$ from the main halo. As an example, Fig.\,\ref{merger} shows the merger of two haloes with masses $M_1 = 0.857 \times 10^{14}\, h^{-1}{\rm M}_{\odot}$ and $M_2 = 0.774 \times 10^{14}\, h^{-1}{\rm M}_{\odot}$ from redshift $z = 0.41$ onwards. The change in their masses is listed in Table~\ref{mergerMasses}. The size of the spheres corresponds to the $r_{200}$  value estimated by the mass of the subhalo via

\begin{equation}
r_{200} = \left(\frac{3}{4\pi}\frac{M}{200\rho_{\mathrm{crit}}(z)}\right)^{1/3} \ ,
\label{eq:r200}
\end{equation}
with $M$ being the mass of the subhalo and $\rho_{\mathrm{crit}}(z)$ the critical density at $z=0.28$. As investigated in detail in \cite{Muldrew2011} and \cite{Behroozi2015}, the {\tt SUBFIND} algorithm has problems resolving infalling substructures as they get closer to the centre of the main halo. The reason for this behaviour is that {\tt SUBFIND} recognises substructures by the presence of saddle points in the density gradient. As the infalling subhalo gets closer to the central halo, it reaches denser regions. The decreasing density contrast leads to problems identifying the infalling subhalo and underpredicts its mass. \cite{Muldrew2011} state that such underprediction could reach as much as 25\% at the virial radius. 

In contrast, the adaptive mesh algorithm {\tt AHF} \citep[][]{Knollmann2009} or the hierarchically code {\tt HBT} \citep[][]{Han2012}, preserve the infalling subhalo further into the central regions. The hierarchical approach of {\tt HBT} is based on linking subhaloes from snapshot to snapshot by tracking the particles of each subhalo and finding the host halo of the progenitor particles.
Despite that, these codes also have their disadvantages. {\tt AHF} assumes a spherical halo which fails when the halo gets elongated by the tidal field. This leads to retaining particles in the halo for too long and underpredicting tidal stripping. Hierarchical codes like {\tt HBT} tend to keep the infalling halo separated even while it reaches the centre and both haloes finally merge.

To investigate the effect of tidal stripping, we trace back all subhaloes of FoF-haloes with suitable mass and evaluate the highest mass each subhalo has before the infall. This is an upper estimate, since a limited amount of tidal stripping is indeed expected. 
However, while taking this effect into account, we still cannot find systems comparable to Abell 2744.
We identified two haloes closest to Abell 2744's case. In the first one, four subhaloes with masses above $10^14 {\rm M}_\odot$ are found in a radius of 1~Mpc around the centre of mass. In the second halo, we found three subhaloes with the same characteristics. All other haloes only contain a central halo with at best one additional subhalo with a mass above $10^{14} {\rm M}_\odot$ before the infall.

Another numerical effect is caused by the way {\tt SUBFIND} assigns particles to subhaloes. Any particle within the FoF group that is not gravitationally bound to a subhalo gets assigned to the central subhalo. Hence, the central subhalo in MXXL would be considerably more massive than the observed one, since the mass of all dark matter that is smoothly distributed between the subhaloes is added to the mass of the central subhalo. We therefore allowed the central halo to be as massive as Abell 2744.

The lack of similar systems in MXXL, one of the largest volume simulation so far, implies that Abell 2744 is one of the most extreme cluster to date, however our analysis does not allow us to conclude definitively on the consistency of Abell 2744 within the $\Lambda$CDM framework. 
More work on the simulation side is required to overcome the caveats highlighted in our discussion, and thus investigate Abell 2744 case in more details.
\begin{table}
\centering
    \begin{tabular}{cccc}
  \hline\hline
    \emph{ID} & $M_{150}$ ($10^{13} {\rm M}_\odot$) & $M_{250}$ ($10^{13} {\rm M}_\odot$)  & $M_{\tt SUBFIND}$ ($10^{13} {\rm M}_\odot$) \\ \hline 
    \emph{Core} & $13.55\pm 0.09$\,\, & $27.7 \pm 0.1$ & $259^{+4}_{-3}$\,\, \\
    \emph{N} & $6.10\pm 0.50$ & $14.7 \pm 0.9$ & $49^{+8}_{-8}$\\
    \emph{NW} & $7.90\pm 0.60$ & $18.0 \pm 1.0$ & \,\,$81^{+14}_{-12}$\\
    $\text{\emph{W}}_\text{\emph{bis}}$ & $5.20\pm 0.60$ & $12.9 \pm 1.1$ & $36^{+8}_{-8}$\\
    \emph{S1} & $5.00\pm 0.40$ & $13.0 \pm 1.0$ & $33^{+5}_{-5}$\\
    \emph{S2} & $5.40\pm 0.50$ & - & $38^{+8}_{-6}$ \\
    \emph{S3} & $6.50\pm 0.60$ & - & \,\,$55^{+11}_{-9}$\\
    \emph{S4} & $5.50\pm 1.20$ & - & \,\,$40^{+19}_{-15}$ \\
    \hline\hline
  \end{tabular}
  \caption{Comparison of mass estimates obtained within an aperture of 150 kpc and 250 kpc and extrapolated mass for {\tt SUBFIND}-haloes for all eight substructures presented in Sect.~\ref{substr_mass}.}
\label{substr_nfw}
\end{table}

\begin{table}
\centering
    \begin{tabular}{ccc}
  \hline\hline
    redshift $z$ & $M_1$ ($10^{14}{\rm M}_{\odot}$)& $M_2$ ($10^{14}{\rm M}_{\odot}$)\\ \hline 
    0.41 & $1.224$ & $1.106$\\
    0.36 & $1.321$ & $0.814$\\
    0.32 & $1.386$ & $0.701$\\
    0.28 & $1.423$ & $0.613$\\
    0.24 & $2.406$ & $0.226$\\
    0.21 & $2.871$ & - \\
    \hline\hline
  \end{tabular}
   \caption{FoF-masses of the two merging subhaloes shown in Fig.\,\ref{merger} at different redshifts.}
\label{mergerMasses}
\end{table}

\subsection{Constraints on Dark Matter's Nature}
While CDM remains the best candidate for dark matter, it is not the only one. We here explore two popular alternatives, warm dark matter (WDM) and self-interacting dark matter (SIDM).

The recent detections of a 3.53 keV emission line in the stacked X-ray spectrum of galaxy clusters \citep{bulbul14} and in individual X-ray spectra of the Perseus cluster and the Andromeda galaxy \citep{boyarsky14}, consistent with the decay of a sterile neutrino with a rest mass of 7.06~keV, has resurrected interest in WDM models. We here focus on results from the COCO simulation \citep[][]{hellwing16,bose16a}, which investigated a cosmology in which dark matter is a 3.3~keV thermal WDM particle.  The primordial power spectrum of this particle closely mimics that of a 7~keV sterile neutrino. While the CDM subhalo mass function continues to rise steeply towards low masses \citep[e.g.][]{jenkins01,tinker08}, the mass function of subhaloes in WDM is heavily suppressed, declining rapidly towards lower masses \citep[M$_{200} < 10^9\, M_\odot$,][]{bose16b}, because of the early free streaming of WDM particles. Above M$_{200} \sim 10^9\, M_\odot$, however, the abundance of subhaloes is nearly identical in WDM and CDM. Since the radial distribution of substructures above $10^9\, M_\odot$ is also similar, we are unable to distinguish between WDM and CDM cosmologies on the mass scales probed by Abell 2744 ($M_{200} > 10^{13}\, M_\odot$).

SIDM was first introduced by \cite{spergel00} as a solution to the missing satellites and core vs.\ cusp problems. If dark-matter particles have a non-zero cross-section for elastic scattering with each other, they are preferentially scattered out of the high-density regions at the centre of dark-matter haloes, thereby leading to lower central densities. Several observational studies have constrained the dark-matter self-interaction cross-section per unit mass, $\sigma / m$, by using galaxy clusters as giant dark-matter particle colliders and looking for a separation between the collisionless galaxies and the (potentially) collisional dark matter, both in major mergers \citep[][]{bradac08a,merten11} and minor mergers \citep[][]{harvey15}. Recently, similar analyses have been performed on individual galaxies moving through galaxy clusters \citep[Abell 3827, ][]{williams11,mohammed14,massey15}, finding potential evidence for non-gravitational dark-matter interactions. 

The substructure within dark-matter haloes offers another potential path toward placing limits on the dark-matter scattering cross-section. As a subhalo moves through the main halo, scattering causes subhalo particles to be flung out of their subhalo, causing a gradual evaporation of the subhalo \citep[][]{gnedin01}. \cite{rocha13} found that, although analytical arguments had overstated the potential for the evaporation of substructures in SIDM collisions, N-body simulations still predict a reduction in the number of subhalos with a given peak circular velocity, and that this reduction is particularly pronounced in the inner regions of the main halo. For dark matter with a velocity-independent cross-section, the effect is larger in cluster-scale haloes, suggesting that measurements of the subhalo mass function within clusters could provide a test of the nature of dark matter.

Following the prescription of \cite{harvey15}, we investigate whether our findings on Abell 2744 favour SIDM over other dark matter candidates. 
Since this test requires sufficiently well constrained positions for all three constituents of a halo, i.e., dark matter, X-ray emitting gas, and galaxies that are halo members, only the \emph{Core} and the \emph{N} halo are sufficiently well measured to yield discriminating constraints on the dark-matter cross-section $\sigma$. 
The position of galaxies within the halos is determined by smoothing their distribution weighted by their individual fluxes. For the \emph{Core} and the \emph{N} halos, we measure ($\alpha = 3.5867571$, $\delta = -30.399759$) and ($\alpha = 3.5781442$, $\delta = -30.357804$) respectively. We adopt the resolution of the grid, $\delta x_{DM} = 16\arcsec$, as a conservative estimate of the error in the weak-lensing position, and assume an uncertainty in both X-ray positions of $\delta x_{\rm gas} = 5\arcsec$. Combining the estimates of the dark-matter cross section obtained from the \emph{N} and \emph{Core} components, we find $\sigma/m = 0.90^{+0.9}_{-0.8}\, $cm$^2$ g$^{-1}$, i.e., support for non-collisional dark matter and thus CDM. We convert this result to an upper limit by computing the one-sided probability and find $\sigma/m{<}1.28\, $cm$^2$ g$^{-1}$(68\% CL). This constraint is tighter than the $\sigma/m{<}3\, $cm$^2$ g$^{-1}$ reported by M11 and is consistent with the results of \cite{harvey15}.

\section{Conclusion}
\label{conclusion}
We present findings from a joint analysis of strong- and weak-gravitational lensing features to reconstruct the mass distribution of Abell 2744 within $4$~Mpc from its BCG, using data obtained with HST and CFHT. Our mass reconstruction requires the presence of eight distinct substructures within $R{<}1$~Mpc, including the cluster main halo, the \emph{Core}, all featuring masses between 0.4 and 1.3${\times}10^{14}$M$_{\odot}$. Complementing our lensing results with insights from deep \emph{Chandra} and \emph{XMM-Newton} observations enables us, in addition, to explore the dynamical status of the cluster.

The main mass concentrations \emph{Core}, \emph{N}, \emph{NW}, previously detected and discussed by M11, are all detected in our analysis. We further detect substructure \emph{W$_{bis}$}, whose location agrees (within the errors) with that of the \emph{W} mass concentration from M16. Like M16, we do not detect any substructure at the location of the feature labelled \emph{W} by M11. Although we here count \emph{W$_{bis}$} as a possible substructure of Abell 2744, we stress that its M/L ratio and the fact that the two spectroscopic redshifts available in this region identify the respective galaxies as background objects strongly suggest that this mass concentration resides behind Abell 2744. Further matching our discoveries with features reported in prior studies, we note that substructure S1 corresponds to the \emph{NE} component of M16. If the shock revealed by \cite{eckert16}, close to the radio relic, is associated with the motion of S1, this substructure would be moving toward the NE direction after first core passage. We also report the new detection of a remnant core, X4, located NE of the cluster core, in the direction of the radio relic, and 40\arcsec away from S1. Substructure S3 is found to match the second component of M11's \emph{NW}, \emph{NW2}. In the same region we also report the newly detected substructure S4, aligned with the axis defined by \emph{NW} and S3, but located farther in the direction of the NW filament reported by \cite{eckert15}. We associate S4 with the X-ray remnant core X2, which was named the \emph{`interloper'} by M11 as they did not find any dark matter counterpart in its vicinity. Finally, the X-ray data allowed us to confirm the presence of a shock SE of the cluster core as previously reported by \cite{owers11}. We also claim the putative detection of another shock, North of the cluster core. If confirmed, this feature would reinforce the scenario proposed by \cite{owers11} of the N--S direction representing the main merger axis.

In the second part of this paper, we search for Abell 2744-like systems in the $\Lambda$CDM simulation MXXL. While clusters of comparable mass are found commonly at similar redshifts ($0.28<z<0.32$), none of them feature halos containing as many sub-halos as Abell 2744, and as close to the centre. Although this discrepancy appears to suggest tension between the results of this work and fundamental predictions of $\Lambda$CDM, we discuss caveats regarding both the simulation and the observational evidence, that render such a conclusion premature. 

Finally, we investigate whether the substructure in Abell 2744 can be used to elucidate the nature of dark matter and, specifically looking at WDM and SIDM, two popular dark-matter candidates. We are unable to draw conclusions regarding WDM since, at the mass range considered here ($0.5-1.3\, 10^{14}$M$_{\odot}$), the WDM and CDM halo mass function are too similar. The situation appears more promising for SIDM whose non-zero self-interaction cross section would lower the post-collision density of subclusters, with the number of mergers being similar to that of CDM. The survival time of SIDM subhalos, however, being shorter than in a CDM universe, fewer substructures are expected in SIDM clusters \cite[as demonstrated by][]{rocha13}, in particular in regions close to the cluster centre. SIDM is not favoured by our findings also because, based on the different position of the light, gas and dark matter in the \emph{Core} and the \emph{N} components, we find no evidence for a non-zero cross section beyond an upper limit of $\sigma_{\rm DM} < 1.28\, $cm$^2$g$^{-1}$(68\% CL), in agreement with \cite{harvey15}. Investigating the low-mass end of the halo mass function, below $10^{9}\, M_{\odot}$, will thus be critical to differentiate between the different dark-matter candidates (see Natarajan et al.\ 2016, \emph{sub.}).

\section*{Acknowledgments}
This work was supported by the Science and Technology Facilities Council [grant number ST/L00075X/1, ST/F001166/1, ST/K501979/1 \& ST/K501979/1] and used the DiRAC Data Centric system at Durham University, operated by the Institute for Computational Cosmology on behalf of the STFC DiRAC HPC Facility (www.dirac.ac.uk [www.dirac.ac.uk]). This equipment was funded by BIS National E-infrastructure capital grant ST/K00042X/1, STFC capital grant ST/H008519/1, and STFC DiRAC Operations grant ST/K003267/1 and Durham University. DiRAC is part of the National E-Infrastructure. 
MJ, DE, and HI thank T. Erben and M. Klein for their help with the WFI data.
MJ, ML, and EJ acknowledge the M\'esocentre d'Aix-Marseille Universit\'e (project number: 14b030).  This study also benefited from the facilities offered by CeSAM (Centre de donn\'eeS Astrophysique de Marseille ({\tt http://lam.oamp.fr/cesam/}). 
MJ thanks St\'ephane Arnouts for his inputs and a fruitful discussion. MJ thanks C\'eline Boehm for carefully reading this paper and sharing her expertise to improve it.
MJ thanks Ian Smail, Alastair Edge, Simon Morris, David Alexander, Richard Bower and Mark Swinbank for a detailed discussion.
JS thanks Matthias Bartelmann for helpful discussions and providing his libastro library for calculating the merger rates.
DH acknowledges the funding support from the Swiss National Science Foundation (SNSF).
CMB acknowledges a Research Fellowship from the Leverhulme Trust.
RM is supported by the Royal Society.
M.S.O acknowledges the funding support from the Australian Research Council through a Future Fellowship (FT140100255).
HYS acknowledges the support from Marie-Curie International Incoming Fellowship (FP7-PEOPLE-2012-IIF/327561) and NSFC of China under grants 11103011.
JPK acknowledges support from the ERC advanced grant LIDA. 
JR acknowledges support from the ERC starting grant CALENDS and the CIG grant 294074. 
ML acknowledges the Centre National de la Recherche Scientifique (CNRS) for its support. 
PN acknowledges support from the National Science Foundation via the grant AST-1044455, AST-1044455, and a theory grant from the Space Telescope Science Institute HST-AR-12144.01-A. 
CT acknowledges the financial support from the Swiss National Science Foundation.

\bibliographystyle{mn2e}
\bibliography{reference}

\begin{thebibliography}{}

\bibitem[\protect\citeauthoryear{Angulo, Springel, White, Jenkins, Baugh \&
  Frenk}{Angulo et~al.}{2012}]{Angulo2012}
Angulo R.~E.,  Springel V.,  White S. D.~M.,  Jenkins A.,  Baugh C.~M.,
  Frenk C.~S.,  2012, Monthly Notices of the Royal Astronomical Society, 426,
  2046

\bibitem[\protect\citeauthoryear{{Atek}, {Richard}, {Kneib}, {Clement},
  {Egami}, {Ebeling}, {Jauzac}, {Jullo}, {Laporte}, {Limousin} \&
  {Natarajan}}{{Atek} et~al.}{2014}]{atek14a}
{Atek} H.,  {Richard} J.,  {Kneib} J.-P.,  {Clement} B.,  {Egami} E.,
  {Ebeling} H.,  {Jauzac} M.,  {Jullo} E.,  {Laporte} N.,  {Limousin} M.,
  {Natarajan} P.,  2014, \apj, 786, 60

\bibitem[\protect\citeauthoryear{{Atek}, {Richard}, {Kneib}, {Jauzac},
  {Schaerer}, {Clement}, {Limousin}, {Jullo}, {Natarajan}, {Egami} \&
  {Ebeling}}{{Atek} et~al.}{2015}]{atek15}
{Atek} H.,  {Richard} J.,  {Kneib} J.-P.,  {Jauzac} M.,  {Schaerer} D.,
  {Clement} B.,  {Limousin} M.,  {Jullo} E.,  {Natarajan} P.,  {Egami} E.,
  {Ebeling} H.,  2015, \apj, 800, 18

\bibitem[\protect\citeauthoryear{{Bacon}, {Massey}, {Refregier} \&
  {Ellis}}{{Bacon} et~al.}{2003}]{bacon03}
{Bacon} D.~J.,  {Massey} R.~J.,  {Refregier} A.~R.,    {Ellis} R.~S.,  2003,
  \mnras, 344, 673

\bibitem[\protect\citeauthoryear{Behroozi, Knebe, Pearce, Elahi, Han, Lux, Mao,
  Muldrew, Potter \& Srisawat}{Behroozi et~al.}{2015}]{Behroozi2015}
Behroozi P.,  Knebe A.,  Pearce F.~R.,  Elahi P.,  Han J.,  Lux H.,  Mao Y.-Y.,
   Muldrew S.~I.,  Potter D.,    Srisawat C.,  2015, \mnras, 454, 3020

\bibitem[\protect\citeauthoryear{{Bertin}}{{Bertin}}{2006}]{bertin06}
{Bertin} E.,  2006, in {Gabriel} C.,  {Arviset} C.,  {Ponz} D.,   {Enrique} S.,
   eds, Astronomical Data Analysis Software and Systems XV Vol.~351 of
  Astronomical Society of the Pacific Conference Series, {Automatic Astrometric
  and Photometric Calibration with SCAMP}.
p.~112

\bibitem[\protect\citeauthoryear{{Bertin}}{{Bertin}}{2010}]{bertin10}
{Bertin} E., , 2010, {SWarp: Resampling and Co-adding FITS Images Together},
  Astrophysics Source Code Library

\bibitem[\protect\citeauthoryear{{Bertin} \& {Arnouts}}{{Bertin} \&
  {Arnouts}}{1996}]{BA96}
{Bertin} E.,  {Arnouts} S.,  1996, \aap, 117, 393

\bibitem[\protect\citeauthoryear{{B{\oe}hm}, {Schewtschenko}, {Wilkinson},
  {Baugh} \& {Pascoli}}{{B{\oe}hm} et~al.}{2014}]{boehm14}
{B{\oe}hm} C.,  {Schewtschenko} J.~A.,  {Wilkinson} R.~J.,  {Baugh} C.~M.,
  {Pascoli} S.,  2014, \mnras, 445, L31

\bibitem[\protect\citeauthoryear{{Bond}, {Kofman} \& {Pogosyan}}{{Bond}
  et~al.}{1996}]{bond96}
{Bond} J.~R.,  {Kofman} L.,    {Pogosyan} D.,  1996, \nat, 380, 603

\bibitem[\protect\citeauthoryear{{Boschin}, {Girardi}, {Spolaor} \&
  {Barrena}}{{Boschin} et~al.}{2006}]{boschin06}
{Boschin} W.,  {Girardi} M.,  {Spolaor} M.,    {Barrena} R.,  2006, \aap, 449,
  461

\bibitem[\protect\citeauthoryear{{Bose}, {Hellwing}, {Frenk}, {Jenkins},
  {Lovell}, {Helly} \& {Li}}{{Bose} et~al.}{2016}]{bose16a}
{Bose} S.,  {Hellwing} W.~A.,  {Frenk} C.~S.,  {Jenkins} A.,  {Lovell} M.~R.,
  {Helly} J.~C.,    {Li} B.,  2016, \mnras, 455, 318

\bibitem[\protect\citeauthoryear{{Bose}, {Hellwing}, {Frenk}, {Jenkins},
  {Lovell}, {Helly}, {Li} \& {Gao}}{{Bose} et~al.}{2016}]{bose16b}
{Bose} S.,  {Hellwing} W.~A.,  {Frenk} C.~S.,  {Jenkins} A.,  {Lovell} M.~R.,
  {Helly} J.~C.,  {Li} B.,    {Gao} L.,  2016, ArXiv e-prints

\bibitem[\protect\citeauthoryear{{Boyarsky}, {Ruchayskiy}, {Iakubovskyi} \&
  {Franse}}{{Boyarsky} et~al.}{2014}]{boyarsky14}
{Boyarsky} A.,  {Ruchayskiy} O.,  {Iakubovskyi} D.,    {Franse} J.,  2014,
  Physical Review Letters, 113, 251301

\bibitem[\protect\citeauthoryear{Boylan-Kolchin, Springel, White, Jenkins \&
  Lemson}{Boylan-Kolchin et~al.}{2009}]{Boylan-Kolchin2009}
Boylan-Kolchin M.,  Springel V.,  White S. D.~M.,  Jenkins A.,    Lemson G.,
  2009, Monthly Notices of the Royal Astronomical Society, 398, 1150

\bibitem[\protect\citeauthoryear{{Bradac}, {Clowe}, {Gonzalez}, {Marshall},
  {Forman}, {Jones}, {Markevitch}, {Randall}, {Schrabback} \&
  {Zaritsky}}{{Bradac} et~al.}{2006}]{bradac06}
{Bradac} M.,  {Clowe} D.,  {Gonzalez} A.~H.,  {Marshall} P.,  {Forman} W.,
  {Jones} C.,  {Markevitch} M.,  {Randall} S.,  {Schrabback} T.,    {Zaritsky}
  D.,  2006, ArXiv Astrophysics e-prints

\bibitem[\protect\citeauthoryear{{Brada{\v c}}, {Schrabback}, {Erben},
  {McCourt}, {Million}, {Mantz}, {Allen}, {Blandford}, {Halkola},
  {Hildebrandt}, {Lombardi}, {Marshall}, {Schneider}, {Treu} \&
  {Kneib}}{{Brada{\v c}} et~al.}{2008}]{bradac08a}
{Brada{\v c}} M.,  {Schrabback} T.,  {Erben} T.,  {McCourt} M.,  {Million} E.,
  {Mantz} A.,  {Allen} S.,  {Blandford} R.,  {Halkola} A.,  {Hildebrandt} H.,
  {Lombardi} M.,  {Marshall} P.,  {Schneider} P.,  {Treu} T.,    {Kneib} J.-P.,
   2008, \apj, 681, 187

\bibitem[\protect\citeauthoryear{{Braglia}, {Pierini} \&
  {B{\"o}hringer}}{{Braglia} et~al.}{2007}]{braglia07}
{Braglia} F.,  {Pierini} D.,    {B{\"o}hringer} H.,  2007, \aap, 470, 425

\bibitem[\protect\citeauthoryear{{Braglia}, {Pierini}, {Biviano} \&
  {B{\"o}hringer}}{{Braglia} et~al.}{2009}]{braglia09}
{Braglia} F.~G.,  {Pierini} D.,  {Biviano} A.,    {B{\"o}hringer} H.,  2009,
  \aap, 500, 947

\bibitem[\protect\citeauthoryear{{Bruzual} \& {Charlot}}{{Bruzual} \&
  {Charlot}}{2003}]{BC03}
{Bruzual} G.,  {Charlot} S.,  2003, \mnras, 344, 1000

\bibitem[\protect\citeauthoryear{{Bulbul}, {Markevitch}, {Foster}, {Smith},
  {Loewenstein} \& {Randall}}{{Bulbul} et~al.}{2014}]{bulbul14}
{Bulbul} E.,  {Markevitch} M.,  {Foster} A.,  {Smith} R.~K.,  {Loewenstein} M.,
     {Randall} S.~W.,  2014, \apj, 789, 13

\bibitem[\protect\citeauthoryear{{Casertano}, {de Mello}, {Dickinson},
  {Ferguson}, {Fruchter}, {Gonzalez-Lopezlira}, {Heyer}, {Hook}, {Levay},
  {Lucas}, {Mack}, {Makidon}, {Mutchler}, {Smith}, {Stiavelli}, {Wiggs} \&
  {Williams}}{{Casertano} et~al.}{2000}]{casertano00}
{Casertano} S.,  {de Mello} D.,  {Dickinson} M.,  {Ferguson} H.~C.,  {Fruchter}
  A.~S.,  {Gonzalez-Lopezlira} R.~A.,  {Heyer} I.,  {Hook} R.~N.,  {Levay} Z.,
  {Lucas} R.~A.,  {Mack} J.,  {Makidon} R.~B.,  {Mutchler} M.,  {Smith} T.~E.,
  {Stiavelli} M.,  {Wiggs} M.~S.,    {Williams} R.~E.,  2000, \aj, 120, 2747

\bibitem[\protect\citeauthoryear{{Castellano}, {Amor{\'{\i}}n}, {Merlin},
  {Fontana}, {McLure}, {M{\'a}rmol-Queralt{\'o}} \& [...]}{{Castellano}
  et~al.}{2016}]{castellano16}
{Castellano} M.,  {Amor{\'{\i}}n} R.,  {Merlin} E.,  {Fontana} A.,  {McLure}
  R.~J.,  {M{\'a}rmol-Queralt{\'o}} E.,    [...] 2016, \aap, 590, A31

\bibitem[\protect\citeauthoryear{{Clowe}, {De Lucia} \& {King}}{{Clowe}
  et~al.}{2004}]{clowe04}
{Clowe} D.,  {De Lucia} G.,    {King} L.,  2004, \mnras, 350, 1038

\bibitem[\protect\citeauthoryear{{Coe}, {Bradley} \& {Zitrin}}{{Coe}
  et~al.}{2015}]{coe15}
{Coe} D.,  {Bradley} L.,    {Zitrin} A.,  2015, \apj, 800, 84

\bibitem[\protect\citeauthoryear{{Coleman}, {Wu} \& {Weedman}}{{Coleman}
  et~al.}{1980}]{CWW}
{Coleman} G.~D.,  {Wu} C.-C.,    {Weedman} D.~W.,  1980, \apjs, 43, 393

\bibitem[\protect\citeauthoryear{{Colless}, {Dalton}, {Maddox}, {Sutherland},
  {Norberg}, {Cole}, {Bland-Hawthorn}, {Bridges}, {Cannon}, {Collins}, {Couch},
  {Cross} \& [...]}{{Colless} et~al.}{2001}]{colless01}
{Colless} M.,  {Dalton} G.,  {Maddox} S.,  {Sutherland} W.,  {Norberg} P.,
  {Cole} S.,  {Bland-Hawthorn} J.,  {Bridges} T.,  {Cannon} R.,  {Collins} C.,
  {Couch} W.,  {Cross} N.,    [...] 2001, \mnras, 328, 1039

\bibitem[\protect\citeauthoryear{{Couch}, {Barger}, {Smail}, {Ellis} \&
  {Sharples}}{{Couch} et~al.}{1998}]{couch98}
{Couch} W.~J.,  {Barger} A.~J.,  {Smail} I.,  {Ellis} R.~S.,    {Sharples}
  R.~M.,  1998, \apj, 497, 188

\bibitem[\protect\citeauthoryear{{Couch} \& {Sharples}}{{Couch} \&
  {Sharples}}{1987}]{couch87}
{Couch} W.~J.,  {Sharples} R.~M.,  1987, \mnras, 229, 423

\bibitem[\protect\citeauthoryear{Davis, Efstathiou, Frenk \& White}{Davis
  et~al.}{1985}]{Davis1985}
Davis M.,  Efstathiou G.,  Frenk C.~S.,    White S. D.~M.,  1985, \aj, 292, 371

\bibitem[\protect\citeauthoryear{{Dietrich}, {Erben}, {Lamer}, {Schneider},
  {Schwope}, {Hartlap} \& {Maturi}}{{Dietrich} et~al.}{2007}]{dietrich07}
{Dietrich} J.~P.,  {Erben} T.,  {Lamer} G.,  {Schneider} P.,  {Schwope} A.,
  {Hartlap} J.,    {Maturi} M.,  2007, \aap, 470, 821

\bibitem[\protect\citeauthoryear{{Ebeling}, {White} \& {Rangarajan}}{{Ebeling}
  et~al.}{2006}]{ebeling06}
{Ebeling} H.,  {White} D.~A.,    {Rangarajan} F.~V.~N.,  2006, \mnras, 368, 65

\bibitem[\protect\citeauthoryear{{Eckert}, {Jauzac}, {Shan}, {Kneib}, {Erben},
  {Israel}, {Jullo}, {Klein}, {Massey}, {Richard} \& {Tchernin}}{{Eckert}
  et~al.}{2015}]{eckert15}
{Eckert} D.,  {Jauzac} M.,  {Shan} H.,  {Kneib} J.-P.,  {Erben} T.,  {Israel}
  H.,  {Jullo} E.,  {Klein} M.,  {Massey} R.,  {Richard} J.,    {Tchernin} C.,
  2015, \nat, 528, 105

\bibitem[\protect\citeauthoryear{{Eckert}, {Jauzac}, {Vazza}, {Owers}, {Kneib},
  {Tchernin}, {Intema} \& {Knowles}}{{Eckert} et~al.}{2016}]{eckert16}
{Eckert} D.,  {Jauzac} M.,  {Vazza} F.,  {Owers} M.,  {Kneib} J.-P.,
  {Tchernin} C.,  {Intema} H.,    {Knowles} K.,  2016, ArXiv e-prints

\bibitem[\protect\citeauthoryear{Eddington}{Eddington}{1913}]{Eddington1913}
Eddington A.~S.,  1913, \mnras, 73, 346

\bibitem[\protect\citeauthoryear{{El{\'{\i}}asd{\'o}ttir}, {Fynbo}, {Hjorth},
  {Ledoux}, {Watson}, {Andersen}, {Malesani}, {Vreeswijk}, {Prochaska},
  {Sollerman} \& {Jaunsen}}{{El{\'{\i}}asd{\'o}ttir}
  et~al.}{2009}]{eliasdottir09}
{El{\'{\i}}asd{\'o}ttir} {\'A}.,  {Fynbo} J.~P.~U.,  {Hjorth} J.,  {Ledoux} C.,
   {Watson} D.~J.,  {Andersen} A.~C.,  {Malesani} D.,  {Vreeswijk} P.~M.,
  {Prochaska} J.~X.,  {Sollerman} J.,    {Jaunsen} A.~O.,  2009, \apj, 697,
  1725

\bibitem[\protect\citeauthoryear{{El{\'{\i}}asd{\'o}ttir}, {Limousin},
  {Richard}, {Hjorth}, {Kneib}, {Natarajan}, {Pedersen}, {Jullo} \&
  {Paraficz}}{{El{\'{\i}}asd{\'o}ttir} et~al.}{2007}]{eliasdottir07}
{El{\'{\i}}asd{\'o}ttir} {\'A}.,  {Limousin} M.,  {Richard} J.,  {Hjorth} J.,
  {Kneib} J.-P.,  {Natarajan} P.,  {Pedersen} K.,  {Jullo} E.,    {Paraficz}
  D.,  2007, ArXiv e-prints, 710

\bibitem[\protect\citeauthoryear{{Erben}, {Hildebrandt}, {Miller}, {van
  Waerbeke}, {Heymans}, {Hoekstra}, {Kitching}, {Mellier}, {Benjamin}, {Blake},
  {Bonnett}, {Cordes} \& [...]}{{Erben} et~al.}{2013}]{erben13}
{Erben} T.,  {Hildebrandt} H.,  {Miller} L.,  {van Waerbeke} L.,  {Heymans} C.,
   {Hoekstra} H.,  {Kitching} T.~D.,  {Mellier} Y.,  {Benjamin} J.,  {Blake}
  C.,  {Bonnett} C.,  {Cordes} O.,    [...] 2013, \mnras, 433, 2545

\bibitem[\protect\citeauthoryear{{Evrard}, {MacFarland}, {Couchman}, {Colberg},
  {Yoshida}, {White}, {Jenkins}, {Frenk}, {Pearce}, {Peacock} \&
  {Thomas}}{{Evrard} et~al.}{2002}]{evrard02}
{Evrard} A.~E.,  {MacFarland} T.~J.,  {Couchman} H.~M.~P.,  {Colberg} J.~M.,
  {Yoshida} N.,  {White} S.~D.~M.,  {Jenkins} A.,  {Frenk} C.~S.,  {Pearce}
  F.~R.,  {Peacock} J.~A.,    {Thomas} P.~A.,  2002, \apj, 573, 7

\bibitem[\protect\citeauthoryear{{Geller} \& {Huchra}}{{Geller} \&
  {Huchra}}{1989}]{GH89}
{Geller} M.~J.,  {Huchra} J.~P.,  1989, Science, 246, 897

\bibitem[\protect\citeauthoryear{{Gilmore} \& {Natarajan}}{{Gilmore} \&
  {Natarajan}}{2009}]{gilmore09}
{Gilmore} J.,  {Natarajan} P.,  2009, \mnras, 396, 354

\bibitem[\protect\citeauthoryear{{Giovannini}, {Tordi} \&
  {Feretti}}{{Giovannini} et~al.}{1999}]{giovannini99}
{Giovannini} G.,  {Tordi} M.,    {Feretti} L.,  1999, \na, 4, 141

\bibitem[\protect\citeauthoryear{{Girardi} \& {Mezzetti}}{{Girardi} \&
  {Mezzetti}}{2001}]{girardi01}
{Girardi} M.,  {Mezzetti} M.,  2001, \apj, 548, 79

\bibitem[\protect\citeauthoryear{{Gnedin} \& {Ostriker}}{{Gnedin} \&
  {Ostriker}}{2001}]{gnedin01}
{Gnedin} O.~Y.,  {Ostriker} J.~P.,  2001, \apj, 561, 61

\bibitem[\protect\citeauthoryear{{Govoni}, {En{\ss}lin}, {Feretti} \&
  {Giovannini}}{{Govoni} et~al.}{2001}]{govoni01a}
{Govoni} F.,  {En{\ss}lin} T.~A.,  {Feretti} L.,    {Giovannini} G.,  2001,
  \aap, 369, 441

\bibitem[\protect\citeauthoryear{Han, Jing, Wang \& Wang}{Han
  et~al.}{2012}]{Han2012}
Han J.,  Jing Y.~P.,  Wang H.,    Wang W.,  2012, \mnras, 427, 2437

\bibitem[\protect\citeauthoryear{{Harvey}, {Kneib} \& {Jauzac}}{{Harvey}
  et~al.}{2016}]{harvey16}
{Harvey} D.,  {Kneib} J.~P.,    {Jauzac} M.,  2016, \mnras, 458, 660

\bibitem[\protect\citeauthoryear{{Harvey}, {Massey}, {Kitching}, {Taylor} \&
  {Tittley}}{{Harvey} et~al.}{2015}]{harvey15}
{Harvey} D.,  {Massey} R.,  {Kitching} T.,  {Taylor} A.,    {Tittley} E.,
  2015, Science, 347, 1462

\bibitem[\protect\citeauthoryear{{Hellwing}, {Frenk}, {Cautun}, {Bose},
  {Helly}, {Jenkins}, {Sawala} \& {Cytowski}}{{Hellwing}
  et~al.}{2016}]{hellwing16}
{Hellwing} W.~A.,  {Frenk} C.~S.,  {Cautun} M.,  {Bose} S.,  {Helly} J.,
  {Jenkins} A.,  {Sawala} T.,    {Cytowski} M.,  2016, \mnras, 457, 3492

\bibitem[\protect\citeauthoryear{{Heymans}, {Van Waerbeke}, {Bacon}, {Berge},
  {Bernstein}, {Bertin}, {Bridle}, {Brown}, {Clowe}, {Dahle}, {Erben}, {Gray}
  \& [...]}{{Heymans} et~al.}{2006}]{heymans06}
{Heymans} C.,  {Van Waerbeke} L.,  {Bacon} D.,  {Berge} J.,  {Bernstein} G.,
  {Bertin} E.,  {Bridle} S.,  {Brown} M.~L.,  {Clowe} D.,  {Dahle} H.,  {Erben}
  T.,  {Gray} M.,    [...] 2006, \mnras, 368, 1323

\bibitem[\protect\citeauthoryear{{High}, {Stubbs}, {Rest}, {Stalder} \&
  {Challis}}{{High} et~al.}{2009}]{high09}
{High} F.~W.,  {Stubbs} C.~W.,  {Rest} A.,  {Stalder} B.,    {Challis} P.,
  2009, \aj, 138, 110

\bibitem[\protect\citeauthoryear{{Hoekstra}, {Bartelmann}, {Dahle}, {Israel},
  {Limousin} \& {Meneghetti}}{{Hoekstra} et~al.}{2013}]{hoekstra13}
{Hoekstra} H.,  {Bartelmann} M.,  {Dahle} H.,  {Israel} H.,  {Limousin} M.,
  {Meneghetti} M.,  2013, \ssr, 177, 75

\bibitem[\protect\citeauthoryear{{Hoekstra}, {Franx}, {Kuijken} \&
  {Squires}}{{Hoekstra} et~al.}{1998}]{hoekstra98}
{Hoekstra} H.,  {Franx} M.,  {Kuijken} K.,    {Squires} G.,  1998, \apj, 504,
  636

\bibitem[\protect\citeauthoryear{Hotchkiss}{Hotchkiss}{2011}]{Hotchkiss2011}
Hotchkiss S.,  2011, Journal of Cosmology and Astroparticle Physics, 2011, 004

\bibitem[\protect\citeauthoryear{{Jauzac}, {Jullo}, {Eckert}, {Ebeling},
  {Richard}, {Limousin}, {Atek}, {Kneib}, {Cl{\'e}ment}, {Egami}, {Harvey},
  {Knowles}, {Massey}, {Natarajan}, {Neichel} \& {Rexroth}}{{Jauzac}
  et~al.}{2015}]{jauzac15a}
{Jauzac} M.,  {Jullo} E.,  {Eckert} D.,  {Ebeling} H.,  {Richard} J.,
  {Limousin} M.,  {Atek} H.,  {Kneib} J.-P.,  {Cl{\'e}ment} B.,  {Egami} E.,
  {Harvey} D.,  {Knowles} K.,  {Massey} R.,  {Natarajan} P.,  {Neichel} B.,
  {Rexroth} M.,  2015, \mnras, 446, 4132

\bibitem[\protect\citeauthoryear{{Jauzac}, {Jullo}, {Kneib}, {Ebeling},
  {Leauthaud}, {Ma}, {Limousin}, {Massey} \& {Richard}}{{Jauzac}
  et~al.}{2012}]{jauzac12}
{Jauzac} M.,  {Jullo} E.,  {Kneib} J.-P.,  {Ebeling} H.,  {Leauthaud} A.,  {Ma}
  C.-J.,  {Limousin} M.,  {Massey} R.,    {Richard} J.,  2012, \mnras, 426,
  3369

\bibitem[\protect\citeauthoryear{{Jauzac}, {Richard}, {Jullo}, {Cl{\'e}ment},
  {Limousin}, {Kneib}, {Ebeling}, {Natarajan}, {Rodney}, {Atek}, {Massey},
  {Eckert}, {Egami} \& {Rexroth}}{{Jauzac} et~al.}{2015}]{jauzac15b}
{Jauzac} M.,  {Richard} J.,  {Jullo} E.,  {Cl{\'e}ment} B.,  {Limousin} M.,
  {Kneib} J.-P.,  {Ebeling} H.,  {Natarajan} P.,  {Rodney} S.,  {Atek} H.,
  {Massey} R.,  {Eckert} D.,  {Egami} E.,    {Rexroth} M.,  2015, \mnras, 452,
  1437

\bibitem[\protect\citeauthoryear{{Jenkins}, {Frenk}, {White}, {Colberg},
  {Cole}, {Evrard}, {Couchman} \& {Yoshida}}{{Jenkins}
  et~al.}{2001}]{jenkins01}
{Jenkins} A.,  {Frenk} C.~S.,  {White} S.~D.~M.,  {Colberg} J.~M.,  {Cole} S.,
  {Evrard} A.~E.,  {Couchman} H.~M.~P.,    {Yoshida} N.,  2001, \mnras, 321,
  372

\bibitem[\protect\citeauthoryear{Jiang, Helly, Cole \& Frenk}{Jiang
  et~al.}{2014}]{jiang14}
Jiang L.,  Helly J.~C.,  Cole S.,    Frenk C.~S.,  2014, Monthly Notices of the
  Royal Astronomical Society, 440, 2115

\bibitem[\protect\citeauthoryear{{Johnson}, {Sharon}, {Bayliss}, {Gladders},
  {Coe} \& {Ebeling}}{{Johnson} et~al.}{2014}]{johnson14}
{Johnson} T.~L.,  {Sharon} K.,  {Bayliss} M.~B.,  {Gladders} M.~D.,  {Coe} D.,
    {Ebeling} H.,  2014, \apj, 797, 48

\bibitem[\protect\citeauthoryear{{Jullo} \& {Kneib}}{{Jullo} \&
  {Kneib}}{2009}]{jullo09}
{Jullo} E.,  {Kneib} J.,  2009, \mnras, 395, 1319

\bibitem[\protect\citeauthoryear{{Jullo}, {Kneib}, {Limousin},
  {El{\'{\i}}asd{\'o}ttir}, {Marshall} \& {Verdugo}}{{Jullo}
  et~al.}{2007}]{jullo07}
{Jullo} E.,  {Kneib} J.-P.,  {Limousin} M.,  {El{\'{\i}}asd{\'o}ttir} {\'A}.,
  {Marshall} P.~J.,    {Verdugo} T.,  2007, New Journal of Physics, 9, 447

\bibitem[\protect\citeauthoryear{{Jullo}, {Pires}, {Jauzac} \& {Kneib}}{{Jullo}
  et~al.}{2014}]{jullo14}
{Jullo} E.,  {Pires} S.,  {Jauzac} M.,    {Kneib} J.-P.,  2014, \mnras, 437,
  3969

\bibitem[\protect\citeauthoryear{{Kaiser}, {Squires} \& {Broadhurst}}{{Kaiser}
  et~al.}{1995}]{KSB95}
{Kaiser} N.,  {Squires} G.,    {Broadhurst} T.,  1995, \apj, 449, 460

\bibitem[\protect\citeauthoryear{{Kalberla}, {Burton}, {Hartmann}, {Arnal},
  {Bajaja}, {Morras} \& {P{\"o}ppel}}{{Kalberla} et~al.}{2005}]{kalberla05}
{Kalberla} P.~M.~W.,  {Burton} W.~B.,  {Hartmann} D.,  {Arnal} E.~M.,  {Bajaja}
  E.,  {Morras} R.,    {P{\"o}ppel} W.~G.~L.,  2005, \aap, 440, 775

\bibitem[\protect\citeauthoryear{{Kawamata}, {Ishigaki}, {Shimasaku}, {Oguri}
  \& {Ouchi}}{{Kawamata} et~al.}{2015}]{kawamata15}
{Kawamata} R.,  {Ishigaki} M.,  {Shimasaku} K.,  {Oguri} M.,    {Ouchi} M.,
  2015, \apj, 804, 103

\bibitem[\protect\citeauthoryear{{Kempner} \& {David}}{{Kempner} \&
  {David}}{2004}]{kempner04}
{Kempner} J.~C.,  {David} L.~P.,  2004, \mnras, 349, 385

\bibitem[\protect\citeauthoryear{Kinney, Calzetti, Bohlin, McQuade,
  Storchi-Bergmann \& Schmitt}{Kinney et~al.}{1996}]{kinney96}
Kinney A.~L.,  Calzetti D.,  Bohlin R.,  McQuade K.,  Storchi-Bergmann T.,
  Schmitt H.,  1996, \apj, 467, 38

\bibitem[\protect\citeauthoryear{{Kneib} \& {Natarajan}}{{Kneib} \&
  {Natarajan}}{2011}]{KN11}
{Kneib} J.-P.,  {Natarajan} P.,  2011, \aapr, 19, 47

\bibitem[\protect\citeauthoryear{Knollmann \& Knebe}{Knollmann \&
  Knebe}{2009}]{Knollmann2009}
Knollmann S.~R.,  Knebe A.,  2009, \apjs, 182, 608

\bibitem[\protect\citeauthoryear{Lacey \& Cole}{Lacey \&
  Cole}{1993}]{Lacey1993}
Lacey C.,  Cole S.,  1993, \mnras, 262, 627

\bibitem[\protect\citeauthoryear{{Lam}, {Broadhurst}, {Diego}, {Lim}, {Coe},
  {Ford} \& {Zheng}}{{Lam} et~al.}{2014}]{lam14}
{Lam} D.,  {Broadhurst} T.,  {Diego} J.~M.,  {Lim} J.,  {Coe} D.,  {Ford}
  H.~C.,    {Zheng} W.,  2014, \apj, 797, 98

\bibitem[\protect\citeauthoryear{{Laporte}, {Streblyanska}, {Clement},
  {P{\'e}rez-Fournon}, {Schaerer}, {Atek}, {Boone}, {Kneib}, {Egami},
  {Mart{\'{\i}}nez-Navajas}, {Marques-Chaves}, {Pell{\'o}} \&
  {Richard}}{{Laporte} et~al.}{2014}]{laporte14}
{Laporte} N.,  {Streblyanska} A.,  {Clement} B.,  {P{\'e}rez-Fournon} I.,
  {Schaerer} D.,  {Atek} H.,  {Boone} F.,  {Kneib} J.-P.,  {Egami} E.,
  {Mart{\'{\i}}nez-Navajas} P.,  {Marques-Chaves} R.,  {Pell{\'o}} R.,
  {Richard} J.,  2014, \aap, 562, L8

\bibitem[\protect\citeauthoryear{{Leauthaud}, {Massey}, {Kneib}, {Rhodes},
  {Johnston}, {Capak}, {Heymans}, {Ellis}, {Koekemoer}, {Le F{\`e}vre},
  {Mellier} \& [...]}{{Leauthaud} et~al.}{2007}]{leauthaud07}
{Leauthaud} A.,  {Massey} R.,  {Kneib} J.,  {Rhodes} J.,  {Johnston} D.~E.,
  {Capak} P.,  {Heymans} C.,  {Ellis} R.~S.,  {Koekemoer} A.~M.,  {Le
  F{\`e}vre} O.,  {Mellier} Y.,    [...] 2007, \apjs, 172, 219

\bibitem[\protect\citeauthoryear{{Leccardi} \& {Molendi}}{{Leccardi} \&
  {Molendi}}{2008}]{leccardi08}
{Leccardi} A.,  {Molendi} S.,  2008, \aap, 487, 461

\bibitem[\protect\citeauthoryear{{Lin}, {Mohr}, {Gonzalez} \& {Stanford}}{{Lin}
  et~al.}{2006}]{lin06}
{Lin} Y.-T.,  {Mohr} J.~J.,  {Gonzalez} A.~H.,    {Stanford} S.~A.,  2006,
  \apjl, 650, L99

\bibitem[\protect\citeauthoryear{{Lotz}, {Koekemoer}, {Coe}, {Grogin}, {Capak},
  {Mack}, {Anderson}, {Avila}, {Barker}, {Borncamp}, {Brammer}, {Durbin} \&
  [...]}{{Lotz} et~al.}{2016}]{lotz16}
{Lotz} J.~M.,  {Koekemoer} A.,  {Coe} D.,  {Grogin} N.,  {Capak} P.,  {Mack}
  J.,  {Anderson} J.,  {Avila} R.,  {Barker} E.~A.,  {Borncamp} D.,  {Brammer}
  G.,  {Durbin} M.,    [...] 2016, ArXiv e-prints

\bibitem[\protect\citeauthoryear{{Luppino} \& {Kaiser}}{{Luppino} \&
  {Kaiser}}{1997}]{LK97}
{Luppino} G.~A.,  {Kaiser} N.,  1997, \apj, 475, 20

\bibitem[\protect\citeauthoryear{{Massey}, {Heymans}, {Berg{\'e}}, {Bernstein},
  {Bridle}, {Clowe}, {Dahle}, {Ellis}, {Erben}, {Hetterscheidt}, {High},
  {Hirata} \& [...]}{{Massey} et~al.}{2007}]{massey07}
{Massey} R.,  {Heymans} C.,  {Berg{\'e}} J.,  {Bernstein} G.,  {Bridle} S.,
  {Clowe} D.,  {Dahle} H.,  {Ellis} R.,  {Erben} T.,  {Hetterscheidt} M.,
  {High} F.~W.,  {Hirata} C.,    [...] 2007, \mnras, 376, 13

\bibitem[\protect\citeauthoryear{{Massey}, {Stoughton}, {Leauthaud}, {Rhodes},
  {Koekemoer}, {Ellis} \& {Shaghoulian}}{{Massey} et~al.}{2010}]{massey10}
{Massey} R.,  {Stoughton} C.,  {Leauthaud} A.,  {Rhodes} J.,  {Koekemoer} A.,
  {Ellis} R.,    {Shaghoulian} E.,  2010, \mnras, 401, 371

\bibitem[\protect\citeauthoryear{{Massey}, {Williams}, {Smit}, {Swinbank},
  {Kitching}, {Harvey}, {Jauzac}, {Israel}, {Clowe}, {Edge}, {Hilton}, {Jullo}
  \& [...]}{{Massey} et~al.}{2015}]{massey15}
{Massey} R.,  {Williams} L.,  {Smit} R.,  {Swinbank} M.,  {Kitching} T.~D.,
  {Harvey} D.,  {Jauzac} M.,  {Israel} H.,  {Clowe} D.,  {Edge} A.,  {Hilton}
  M.,  {Jullo} E.,    [...] 2015, \mnras, 449, 3393

\bibitem[\protect\citeauthoryear{{Medezinski}, {Umetsu}, {Okabe}, {Nonino},
  {Molnar}, {Massey}, {Dupke} \& {Merten}}{{Medezinski}
  et~al.}{2016}]{medezinski16}
{Medezinski} E.,  {Umetsu} K.,  {Okabe} N.,  {Nonino} M.,  {Molnar} S.,
  {Massey} R.,  {Dupke} R.,    {Merten} J.,  2016, \apj, 817, 24

\bibitem[\protect\citeauthoryear{{Merlin}, {Amor{\'{\i}}n}, {Castellano},
  {Fontana}, {Buitrago}, {Dunlop}, {Elbaz} \& [...]}{{Merlin}
  et~al.}{2016}]{merlin16}
{Merlin} E.,  {Amor{\'{\i}}n} R.,  {Castellano} M.,  {Fontana} A.,  {Buitrago}
  F.,  {Dunlop} J.~S.,  {Elbaz} D.,    [...] 2016, \aap, 590, A30

\bibitem[\protect\citeauthoryear{{Merten}, {Coe}, {Dupke}, {Massey}, {Zitrin},
  {Cypriano}, {Okabe}, {Frye}, {Braglia}, {Jim{\'e}nez-Teja}, {Ben{\'{\i}}tez}
  \& [...]}{{Merten} et~al.}{2011}]{merten11}
{Merten} J.,  {Coe} D.,  {Dupke} R.,  {Massey} R.,  {Zitrin} A.,  {Cypriano}
  E.~S.,  {Okabe} N.,  {Frye} B.,  {Braglia} F.~G.,  {Jim{\'e}nez-Teja} Y.,
  {Ben{\'{\i}}tez} N.,    [...] 2011, \mnras, 417, 333

\bibitem[\protect\citeauthoryear{{Mohammed}, {Liesenborgs}, {Saha} \&
  {Williams}}{{Mohammed} et~al.}{2014}]{mohammed14}
{Mohammed} I.,  {Liesenborgs} J.,  {Saha} P.,    {Williams} L.~L.~R.,  2014,
  \mnras, 439, 2651

\bibitem[\protect\citeauthoryear{{Montes} \& {Trujillo}}{{Montes} \&
  {Trujillo}}{2014}]{montes14}
{Montes} M.,  {Trujillo} I.,  2014, \apj, 794, 137

\bibitem[\protect\citeauthoryear{Muldrew, Pearce \& Power}{Muldrew
  et~al.}{2011}]{Muldrew2011}
Muldrew S.~I.,  Pearce F.~R.,    Power C.,  2011, \mnras, 410, 2617

\bibitem[\protect\citeauthoryear{{Murray}, {Power} \& {Robotham}}{{Murray}
  et~al.}{2013}]{Murray2013}
{Murray} S.~G.,  {Power} C.,    {Robotham} A.~S.~G.,  2013, Astronomy and
  Computing, 3, 23

\bibitem[\protect\citeauthoryear{{Natarajan} \& {Kneib}}{{Natarajan} \&
  {Kneib}}{1997}]{natarajan97}
{Natarajan} P.,  {Kneib} J.-P.,  1997, \mnras, 287, 833

\bibitem[\protect\citeauthoryear{{Navarro}, {Frenk} \& {White}}{{Navarro}
  et~al.}{1996}]{NFW96}
{Navarro} J.~F.,  {Frenk} C.~S.,    {White} S.~D.~M.,  1996, \apj, 462, 563

\bibitem[\protect\citeauthoryear{{Neto}, {Gao}, {Bett}, {Cole}, {Navarro},
  {Frenk}, {White}, {Springel} \& {Jenkins}}{{Neto} et~al.}{2007}]{neto07}
{Neto} A.~F.,  {Gao} L.,  {Bett} P.,  {Cole} S.,  {Navarro} J.~F.,  {Frenk}
  C.~S.,  {White} S.~D.~M.,  {Springel} V.,    {Jenkins} A.,  2007, \mnras,
  381, 1450

\bibitem[\protect\citeauthoryear{{Ogrean}, {van Weeren}, {Jones}, {Clarke},
  {Sayers}, {Mroczkowski}, {Nulsen}, {Forman}, {Murray}, {Pandey-Pommier} \&
  [...]}{{Ogrean} et~al.}{2015}]{ogrean15}
{Ogrean} G.~A.,  {van Weeren} R.~J.,  {Jones} C.,  {Clarke} T.~E.,  {Sayers}
  J.,  {Mroczkowski} T.,  {Nulsen} P.~E.~J.,  {Forman} W.,  {Murray} S.~S.,
  {Pandey-Pommier} M.,    [...] 2015, \apj, 812, 153

\bibitem[\protect\citeauthoryear{{Orr{\'u}}, {Murgia}, {Feretti}, {Govoni},
  {Brunetti}, {Giovannini}, {Girardi} \& {Setti}}{{Orr{\'u}}
  et~al.}{2007}]{orru07}
{Orr{\'u}} E.,  {Murgia} M.,  {Feretti} L.,  {Govoni} F.,  {Brunetti} G.,
  {Giovannini} G.,  {Girardi} M.,    {Setti} G.,  2007, \aap, 467, 943

\bibitem[\protect\citeauthoryear{{Owers}, {Randall}, {Nulsen}, {Couch}, {David}
  \& {Kempner}}{{Owers} et~al.}{2011}]{owers11}
{Owers} M.~S.,  {Randall} S.~W.,  {Nulsen} P.~E.~J.,  {Couch} W.~J.,  {David}
  L.~P.,    {Kempner} J.~C.,  2011, \apj, 728, 27

\bibitem[\protect\citeauthoryear{{Pearce}, {Van Weeren}, {Jones}, {Forman},
  {Ogrean}, {Andrade-Santos}, {Kraft}, {Dawson}, {Br{\"u}ggen}, {Roediger},
  {Bulbul} \& {Mroczkowski}}{{Pearce} et~al.}{2016}]{pearce16}
{Pearce} C.,  {Van Weeren} R.~J.,  {Jones} C.,  {Forman} W.~R.,  {Ogrean}
  G.~A.,  {Andrade-Santos} F.,  {Kraft} R.~P.,  {Dawson} W.,  {Br{\"u}ggen} M.,
   {Roediger} E.,  {Bulbul} E.,    {Mroczkowski} T.,  2016, in American
  Astronomical Society Meeting Abstracts Vol.~227 of American Astronomical
  Society Meeting Abstracts, {A Cosmic Train Wreck: JVLA Radio Observations of
  the HST Frontier Fields Cluster Abell 2744}.
p. 235.03

\bibitem[\protect\citeauthoryear{{Planck Collaboration}, Ade, Aghanim, Arnaud,
  Ashdown, Aumont, Baccigalupi, Banday, Barreiro, Bartolo, Battaner, Battye,
  Benabed, Beno{\^{i}}t \& [...]}{{Planck Collaboration}
  et~al.}{2015}]{Planck2015}
{Planck Collaboration} Ade P. a.~R.,  Aghanim N.,  Arnaud M.,  Ashdown M.,
  Aumont J.,  Baccigalupi C.,  Banday a.~J.,  Barreiro R.~B.,  Bartolo N.,
  Battaner E.,  Battye R.,  Benabed K.,  Beno{\^{i}}t A.,    [...] 2015, arXiv
  preprint, 1502.01589

\bibitem[\protect\citeauthoryear{{Rawle}, {Altieri}, {Egami},
  {P{\'e}rez-Gonz{\'a}lez}, {Richard}, {Santos}, {Valtchanov}, {Walth}, {Bouy},
  {Haines} \& {Okabe}}{{Rawle} et~al.}{2014}]{rawle14}
{Rawle} T.~D.,  {Altieri} B.,  {Egami} E.,  {P{\'e}rez-Gonz{\'a}lez} P.~G.,
  {Richard} J.,  {Santos} J.~S.,  {Valtchanov} I.,  {Walth} G.,  {Bouy} H.,
  {Haines} C.~P.,    {Okabe} N.,  2014, \mnras, 442, 196

\bibitem[\protect\citeauthoryear{{Rhodes}, {Refregier} \& {Groth}}{{Rhodes}
  et~al.}{2000}]{rhodes00}
{Rhodes} J.,  {Refregier} A.,    {Groth} E.~J.,  2000, \apj, 536, 79

\bibitem[\protect\citeauthoryear{{Rhodes}, {Massey}, {Albert}, {Collins},
  {Ellis}, {Heymans}, {Gardner}, {Kneib}, {Koekemoer}, {Leauthaud}, {Mellier},
  {Refregier}, {Taylor} \& {Van Waerbeke}}{{Rhodes} et~al.}{2007}]{rhodes07}
{Rhodes} J.~D.,  {Massey} R.~J.,  {Albert} J.,  {Collins} N.,  {Ellis} R.~S.,
  {Heymans} C.,  {Gardner} J.~P.,  {Kneib} J.-P.,  {Koekemoer} A.,  {Leauthaud}
  A.,  {Mellier} Y.,  {Refregier} A.,  {Taylor} J.~E.,    {Van Waerbeke} L.,
  2007, \apjs, 172, 203

\bibitem[\protect\citeauthoryear{{Richard}, {Jauzac}, {Limousin}, {Jullo},
  {Cl{\'e}ment}, {Ebeling}, {Kneib}, {Atek}, {Natarajan}, {Egami}, {Livermore}
  \& {Bower}}{{Richard} et~al.}{2014}]{richard14}
{Richard} J.,  {Jauzac} M.,  {Limousin} M.,  {Jullo} E.,  {Cl{\'e}ment} B.,
  {Ebeling} H.,  {Kneib} J.-P.,  {Atek} H.,  {Natarajan} P.,  {Egami} E.,
  {Livermore} R.,    {Bower} R.,  2014, \mnras, 444, 268

\bibitem[\protect\citeauthoryear{{Rix}, {Barden}, {Beckwith}, {Bell}, {Borch},
  {Caldwell}, {H{\"a}ussler}, {Jahnke}, {Jogee}, {McIntosh}, {Meisenheimer},
  {Peng}, {Sanchez}, {Somerville}, {Wisotzki} \& {Wolf}}{{Rix}
  et~al.}{2004}]{rix04}
{Rix} H.-W.,  {Barden} M.,  {Beckwith} S.~V.~W.,  {Bell} E.~F.,  {Borch} A.,
  {Caldwell} J.~A.~R.,  {H{\"a}ussler} B.,  {Jahnke} K.,  {Jogee} S.,
  {McIntosh} D.~H.,  {Meisenheimer} K.,  {Peng} C.~Y.,  {Sanchez} S.~F.,
  {Somerville} R.~S.,  {Wisotzki} L.,    {Wolf} C.,  2004, \apjs, 152, 163

\bibitem[\protect\citeauthoryear{{Rocha}, {Peter}, {Bullock}, {Kaplinghat},
  {Garrison-Kimmel}, {O{\~n}orbe} \& {Moustakas}}{{Rocha}
  et~al.}{2013}]{rocha13}
{Rocha} M.,  {Peter} A.~H.~G.,  {Bullock} J.~S.,  {Kaplinghat} M.,
  {Garrison-Kimmel} S.,  {O{\~n}orbe} J.,    {Moustakas} L.~A.,  2013, \mnras,
  430, 81

\bibitem[\protect\citeauthoryear{{Saunders}, {Bridges}, {Gillingham}, {Haynes},
  {Smith}, {Whittard}, {Churilov}, {Lankshear}, {Croom}, {Jones} \&
  {Boshuizen}}{{Saunders} et~al.}{2004}]{saunders04}
{Saunders} W.,  {Bridges} T.,  {Gillingham} P.,  {Haynes} R.,  {Smith} G.~A.,
  {Whittard} J.~D.,  {Churilov} V.,  {Lankshear} A.,  {Croom} S.,  {Jones} D.,
    {Boshuizen} C.,  2004, in {Moorwood} A.~F.~M.,  {Iye} M.,  eds,
  Ground-based Instrumentation for Astronomy Vol.~5492 of \procspie, {AAOmega:
  a scientific and optical overview}.
pp 389--400

\bibitem[\protect\citeauthoryear{{Schaye}, {Crain}, {Bower}, {Furlong},
  {Schaller}, {Theuns}, {Dalla Vecchia}, {Frenk}, {McCarthy}, {Helly} \&
  [...]}{{Schaye} et~al.}{2015}]{schaye15}
{Schaye} J.,  {Crain} R.~A.,  {Bower} R.~G.,  {Furlong} M.,  {Schaller} M.,
  {Theuns} T.,  {Dalla Vecchia} C.,  {Frenk} C.~S.,  {McCarthy} I.~G.,  {Helly}
  J.~C.,    [...] 2015, \mnras, 446, 521

\bibitem[\protect\citeauthoryear{{Schirmer}, {Carrasco}, {Pessev}, {Garrel},
  {Winge}, {Neichel} \& {Vidal}}{{Schirmer} et~al.}{2015}]{schirmer15}
{Schirmer} M.,  {Carrasco} E.~R.,  {Pessev} P.,  {Garrel} V.,  {Winge} C.,
  {Neichel} B.,    {Vidal} F.,  2015, \apjs, 217, 33

\bibitem[\protect\citeauthoryear{{Schmidt}, {Treu}, {Brammer}, {Brada{\v c}},
  {Wang}, {Dijkstra}, {Dressler}, {Fontana}, {Gavazzi}, {Henry}, {Hoag},
  {Jones}, {Kelly}, {Malkan}, {Mason}, {Pentericci} \& [...]}{{Schmidt}
  et~al.}{2014}]{schmidt14}
{Schmidt} K.~B.,  {Treu} T.,  {Brammer} G.~B.,  {Brada{\v c}} M.,  {Wang} X.,
  {Dijkstra} M.,  {Dressler} A.,  {Fontana} A.,  {Gavazzi} R.,  {Henry} A.~L.,
  {Hoag} A.,  {Jones} T.~A.,  {Kelly} P.~L.,  {Malkan} M.~A.,  {Mason} C.,
  {Pentericci} L.,    [...] 2014, \apjl, 782, L36

\bibitem[\protect\citeauthoryear{{Shan}, {Kneib}, {Tao}, {Fan}, {Jauzac},
  {Limousin}, {Massey}, {Rhodes}, {Thanjavur} \& {McCracken}}{{Shan}
  et~al.}{2012}]{shan12}
{Shan} H.,  {Kneib} J.-P.,  {Tao} C.,  {Fan} Z.,  {Jauzac} M.,  {Limousin} M.,
  {Massey} R.,  {Rhodes} J.,  {Thanjavur} K.,    {McCracken} H.~J.,  2012,
  \apj, 748, 56

\bibitem[\protect\citeauthoryear{{Sharp}, {Saunders}, {Smith}, {Churilov},
  {Correll}, {Dawson}, {Farrel}, {Frost}, {Haynes}, {Heald}, {Lankshear},
  {Mayfield}, {Waller} \& {Whittard}}{{Sharp} et~al.}{2006}]{sharp06}
{Sharp} R.,  {Saunders} W.,  {Smith} G.,  {Churilov} V.,  {Correll} D.,
  {Dawson} J.,  {Farrel} T.,  {Frost} G.,  {Haynes} R.,  {Heald} R.,
  {Lankshear} A.,  {Mayfield} D.,  {Waller} L.,    {Whittard} D.,  2006, in
  Society of Photo-Optical Instrumentation Engineers (SPIE) Conference Series
  Vol.~6269 of \procspie, {Performance of AAOmega: the AAT multi-purpose
  fiber-fed spectrograph}.
p. 62690G

\bibitem[\protect\citeauthoryear{{Skilling}}{{Skilling}}{1998}]{skilling98}
{Skilling} J.,  1998, in {G.~J.~Erickson, J.~T.~Rychert, \& C.~R.~Smith} ed.,
  Maximum Entropy and Bayesian Methods {Massive Inference and Maximum Entropy}.
pp~1--+

\bibitem[\protect\citeauthoryear{{Smith}, {Saunders}, {Bridges}, {Churilov},
  {Lankshear}, {Dawson}, {Correll}, {Waller}, {Haynes} \& {Frost}}{{Smith}
  et~al.}{2004}]{smith04}
{Smith} G.~A.,  {Saunders} W.,  {Bridges} T.,  {Churilov} V.,  {Lankshear} A.,
  {Dawson} J.,  {Correll} D.,  {Waller} L.,  {Haynes} R.,    {Frost} G.,  2004,
  in {Moorwood} A.~F.~M.,  {Iye} M.,  eds, Ground-based Instrumentation for
  Astronomy Vol.~5492 of \procspie, {AAOmega: a multipurpose fiber-fed
  spectrograph for the AAT}.
pp 410--420

\bibitem[\protect\citeauthoryear{{Smith}, {Brickhouse}, {Liedahl} \&
  {Raymond}}{{Smith} et~al.}{2001}]{smith01xray}
{Smith} R.~K.,  {Brickhouse} N.~S.,  {Liedahl} D.~A.,    {Raymond} J.~C.,
  2001, \apjl, 556, L91

\bibitem[\protect\citeauthoryear{{Snowden}, {Mushotzky}, {Kuntz} \&
  {Davis}}{{Snowden} et~al.}{2008}]{snowden08}
{Snowden} S.~L.,  {Mushotzky} R.~F.,  {Kuntz} K.~D.,    {Davis} D.~S.,  2008,
  \aap, 478, 615

\bibitem[\protect\citeauthoryear{{Spergel} \& {Steinhardt}}{{Spergel} \&
  {Steinhardt}}{2000}]{spergel00}
{Spergel} D.~N.,  {Steinhardt} P.~J.,  2000, Physical Review Letters, 84, 3760

\bibitem[\protect\citeauthoryear{{Springel}, {Frenk} \& {White}}{{Springel}
  et~al.}{2006}]{springel06}
{Springel} V.,  {Frenk} C.~S.,    {White} S.~D.~M.,  2006, \nat, 440, 1137

\bibitem[\protect\citeauthoryear{{Springel}, {White}, {Jenkins}, {Frenk},
  {Yoshida}, {Gao}, {Navarro}, {Thacker} \& [...]}{{Springel}
  et~al.}{2005}]{springel05}
{Springel} V.,  {White} S.~D.~M.,  {Jenkins} A.,  {Frenk} C.~S.,  {Yoshida} N.,
   {Gao} L.,  {Navarro} J.,  {Thacker} R.,    [...] 2005, \nat, 435, 629

\bibitem[\protect\citeauthoryear{Springel, White, Jenkins, Frenk, Yoshida, Gao,
  Navarro, Thacker, Croton, Helly, Peacock, Cole, Thomas, Couchman, Evrard,
  Colberg \& Pearce}{Springel et~al.}{2005}]{Springel2005}
Springel V.,  White S. D.~M.,  Jenkins A.,  Frenk C.~S.,  Yoshida N.,  Gao L.,
  Navarro J.,  Thacker R.,  Croton D.,  Helly J.,  Peacock J.~A.,  Cole S.,
  Thomas P.,  Couchman H.,  Evrard A.,  Colberg J.,    Pearce F.,  2005,
  Nature, 435, 629

\bibitem[\protect\citeauthoryear{Springel, White, Tormen \& Kauffmann}{Springel
  et~al.}{2001}]{Springel2001}
Springel V.,  White S. D.~M.,  Tormen G.,    Kauffmann G.,  2001, Monthly
  Notices of the Royal Astronomical Society, 328, 726

\bibitem[\protect\citeauthoryear{{Tinker}, {Kravtsov}, {Klypin}, {Abazajian},
  {Warren}, {Yepes}, {Gottl{\"o}ber} \& {Holz}}{{Tinker}
  et~al.}{2008}]{tinker08}
{Tinker} J.,  {Kravtsov} A.~V.,  {Klypin} A.,  {Abazajian} K.,  {Warren} M.,
  {Yepes} G.,  {Gottl{\"o}ber} S.,    {Holz} D.~E.,  2008, \apj, 688, 709

\bibitem[\protect\citeauthoryear{{Treu}, {Schmidt}, {Brammer}, {Vulcani},
  {Wang}, {Brada{\v c}}, {Dijkstra}, {Dressler}, {Fontana}, {Gavazzi}, {Henry},
  {Hoag} \& [...]}{{Treu} et~al.}{2015}]{treu15}
{Treu} T.,  {Schmidt} K.~B.,  {Brammer} G.~B.,  {Vulcani} B.,  {Wang} X.,
  {Brada{\v c}} M.,  {Dijkstra} M.,  {Dressler} A.,  {Fontana} A.,  {Gavazzi}
  R.,  {Henry} A.~L.,  {Hoag} A.,    [...] 2015, \apj, 812, 114

\bibitem[\protect\citeauthoryear{{Viel}, {Haehnelt} \& {Springel}}{{Viel}
  et~al.}{2010}]{viel11}
{Viel} M.,  {Haehnelt} M.~G.,    {Springel} V.,  2010, \jcap, 6, 015

\bibitem[\protect\citeauthoryear{{Vogelsberger}, {Genel}, {Springel}, {Torrey},
  {Sijacki}, {Xu}, {Snyder}, {Nelson} \& {Hernquist}}{{Vogelsberger}
  et~al.}{2014}]{vogelsberger14}
{Vogelsberger} M.,  {Genel} S.,  {Springel} V.,  {Torrey} P.,  {Sijacki} D.,
  {Xu} D.,  {Snyder} G.,  {Nelson} D.,    {Hernquist} L.,  2014, \mnras, 444,
  1518

\bibitem[\protect\citeauthoryear{Waizmann, Ettori \& Bartelmann}{Waizmann
  et~al.}{2013}]{Waizmann2013}
Waizmann J.-C.,  Ettori S.,    Bartelmann M.,  2013, Monthly Notices of the
  Royal Astronomical Society, 432, 914

\bibitem[\protect\citeauthoryear{{Wang}, {Hoag}, {Huang}, {Treu}, {Brada{\v
  c}}, {Schmidt}, {Brammer} \& [...]}{{Wang} et~al.}{2015}]{wang15}
{Wang} X.,  {Hoag} A.,  {Huang} K.-H.,  {Treu} T.,  {Brada{\v c}} M.,
  {Schmidt} K.~B.,  {Brammer} G.~B.,    [...] 2015, \apj, 811, 29

\bibitem[\protect\citeauthoryear{{Williams} \& {Saha}}{{Williams} \&
  {Saha}}{2011}]{williams11}
{Williams} L.~L.~R.,  {Saha} P.,  2011, \mnras, 415, 448

\bibitem[\protect\citeauthoryear{{Zhang}, {Finoguenov}, {B{\"o}hringer},
  {Ikebe}, {Matsushita} \& {Schuecker}}{{Zhang} et~al.}{2004}]{zhang04}
{Zhang} Y.-Y.,  {Finoguenov} A.,  {B{\"o}hringer} H.,  {Ikebe} Y.,
  {Matsushita} K.,    {Schuecker} P.,  2004, \aap, 413, 49

\bibitem[\protect\citeauthoryear{{Zheng}, {Shu}, {Moustakas}, {Zitrin}, {Ford},
  {Huang}, {Broadhurst}, {Molino}, {Diego}, {Infante}, {Bauer}, {Kelson} \&
  {Smit}}{{Zheng} et~al.}{2014}]{zheng14}
{Zheng} W.,  {Shu} X.,  {Moustakas} J.,  {Zitrin} A.,  {Ford} H.~C.,  {Huang}
  X.,  {Broadhurst} T.,  {Molino} A.,  {Diego} J.~M.,  {Infante} L.,  {Bauer}
  F.~E.,  {Kelson} D.~D.,    {Smit} R.,  2014, \apj, 795, 93

\bibitem[\protect\citeauthoryear{{Zitrin}, {Ellis}, {Belli} \&
  {Stark}}{{Zitrin} et~al.}{2015}]{zitrin15}
{Zitrin} A.,  {Ellis} R.~S.,  {Belli} S.,    {Stark} D.~P.,  2015, \apjl, 805,
  L7

\bibitem[\protect\citeauthoryear{{Zitrin}, {Zheng}, {Broadhurst}, {Moustakas},
  {Lam}, {Shu}, {Huang}, {Diego}, {Ford}, {Lim}, {Bauer}, {Infante}, {Kelson}
  \& {Molino}}{{Zitrin} et~al.}{2014}]{zitrin14}
{Zitrin} A.,  {Zheng} W.,  {Broadhurst} T.,  {Moustakas} J.,  {Lam} D.,  {Shu}
  X.,  {Huang} X.,  {Diego} J.~M.,  {Ford} H.,  {Lim} J.,  {Bauer} F.~E.,
  {Infante} L.,  {Kelson} D.~D.,    {Molino} A.,  2014, \apjl, 793, L12

\end{thebibliography}


\label{lastpage}

\end{document}